\documentclass[10pt,aps,prx,preprintnumbers,superscriptaddress,showpacs,
twocolumn,floatfix,nofootinbib]{revtex4-2}
\usepackage{graphicx}
\usepackage{amsmath}
\usepackage{slashed}
\usepackage{bm}

\def\d{\partial}
\newcommand{\bfm}[1]{\mbox{\boldmath$#1$}}

\newcommand{\gsim}{\;\rlap{\lower 3.5 pt \hbox{$\mathchar \sim$}} \raise 1pt
\hbox {$>$}\;}
\newcommand{\lsim}{\;\rlap{\lower 3.5 pt \hbox{$\mathchar \sim$}} \raise 1pt
\hbox {$<$}\;}

\begin{document}

\title{\boldmath
Nonlinear Dynamics of Rapidly Driven Systems
\unboldmath}
\author{Afshin Besharat}
\email[]{abeshara@ualberta.ca}
\affiliation{Department of Physics, University of Alberta, Edmonton, Alberta T6G
2J1, Canada}
\author{Alexander A. Penin}
\email[]{penin@ualberta.ca}
\affiliation{Department of Physics, University of Alberta, Edmonton, Alberta T6G
2J1, Canada}

\begin{abstract}

We consider systems characterized by the presence of a rapidly
oscillating force. A general method is presented for the
construction of the effective action governing the large-scale
nonlinear dynamics of such systems order by order in inverse
powers of the oscillation frequency $\omega$. The explicit
expression for the effective Lagrangian is derived up to ${\cal
O}(1/\omega^6)$ next-to-next-to-leading approximation in the
strong-drive regime. The general structure of the
high-frequency expansion reveals a broad class of nonlinear
systems whose transition curves are identical to those of the
linear Mathieu equation, which enables a fully nonperturbative
stability analysis in the case of  strong driving and
nonlinearity. The method is generalized to velocity-dependent
forces and configuration space with curvature, characteristic
to  systems with constraints. Several applications are
discussed in detail, including the dynamical magnetic trapping
of electric charges.
\end{abstract}

\maketitle

\section{Introduction}
\label{sec::intro}

Since the classical work \cite{Kapitza:1951} the dynamics of
particles in a rapidly oscillating field has been studied in a
wide range of problems  from dynamical chaos
\cite{Moon:1985,Friedman:2001} to quantum computing
\cite{Blatt:2008,Monroe:2021} and Floquet engineering of
quantum materials \cite{Oka:2019,Wintersperger:2020rqb} with
the renowned application in the design of the ion traps
\cite{Paul:1990} as well as the neutral atom traps
\cite{Cornell:1991,Jiang:2023}.  Theoretical description of
this class of systems is based on the concept of averaging,
when the effect of the oscillating field is smeared out and the
long-time evolution is governed by the resulting effective
interaction naturally obtained  as a series in the ratio of the
oscillation period to a characteristic time scale of the
averaged system. However, most of the applications so far deal
with the leading order approximation in the high-frequency
limit. The role of the high-order and nonlinear effects remains
a fundamental open question in the theory of dynamical
stabilization. Given the long history and the  importance of
the problem,  surprisingly little has been known about the
high-order  behavior of the generic nonlinear three-dimensional
systems, despite numerous studies performed within different
frameworks (see \cite{Fidlin:2000,Rahav:2003a,Rahav:2003b,
Blekhman:2016,Kremer:2016,Venkatraman:2021lwv,Maggia:2020,
Thomsen:2023,Beneke:2023ndu,Sahoo:2024} and references
therein). The main challenge of the perturbative analysis is in
finding an algorithm of a systematic {\it high-frequency
expansion} for the time-averaged effective action, which
renders high-order calculations  feasible. Such an algorithm
has been recently introduced in a letter \cite{Penin:2023dtn},
where the explicit next-to-leading result has been obtained for
the three-dimensional systems, and the analysis has been forced
through the next-to-next-to-leading order in the case of one
dimension. A remarkable finding of \cite{Penin:2023dtn} is that
beyond the leading order the systems reveal highly nontrivial
nonlinear behavior - the trajectories of the particles in the
rapidly oscillating fields coincide with the geodesics in a
Riemann space with the curvature determined by the spatial
distribution of the oscillating field amplitude, {\it i.e.} the
systems emulate post-Newtonian general relativity.

In this paper we give a detailed account of the method
\cite{Penin:2023dtn}, present the calculation of the effective
action in any number of spatial dimensions up to the
next-to-next-to-leading order, and describe the generalization
of the method to the velocity-dependent forces and to the
configuration space with curvature, naturally appearing in the
mechanical systems with constraints.  A general criterion is
formulated determining when the stability analysis of a rapidly
driven nonlinear system can be reduced to that of the linear
Mathieu equation to all orders of the high-frequency expansion,
thereby extending the results beyond the perturbation theory.
We present explicit solutions for a number of  mechanical
systems including the classical problem of Kapitza pendulum.
Finally, the method is applied to the charged particle dynamics
in the inhomogeneous rapidly oscillating magnetic field, which
sets up the foundation for the new type of dynamical magnetic
traps \cite{Besharat:2025vbw}.

\vspace*{-1mm}

\section{General formalism}
\label{sec::1}

Our  starting point is the classical equation of motion for a
particle of mass $m$ subjected to a static force $-\bfm
G$ and a periodic force $-{\bfm F}\cos \omega t$
\begin{equation}
m\ddot{\bfm R}+\bfm G(\bfm R)+\bfm F(\bfm R)\cos{\omega t}=0\,,
\label{eq::EOMfull}
\end{equation}
where the dot stands for the time derivative $d/dt$ and the
bold fonts indicate  $d$-dimensional vectors for any $d$. The
periodic drive is limited to a single harmonic for the clarity
of the presentation but the inclusion of higher harmonics is
rather straightforward. We do not specify the nature of the
external fields to keep the discussion general and  consider
the limit of fast oscillation. Let us quantify this condition
as it plays a crucial role for the determination of the
expansion parameter and the power counting rules. For a system
of a characteristic size $L$ a typical velocity acquired by the
particle   under the action of the time-independent force is
${v}\sim \left( {G} L/m\right)^{1/2}$. One can define a
``reference'' velocity $c=L\omega$ and the oscillations are
considered fast when $v/c\ll 1$. Our approach  relies on the
effective theory concept to separate the ``slow'' large-scale
dynamics characterized by the velocity $v$ from the ``fast''
small-scale dynamics characterized by the velocity $c$ and
manifested through the power corrections in the scale ratio to
the effective action. It is convenient to introduce the
dimensionless variables $\omega t\to t$, $R/L\to R$ so that the
equation of motion becomes
\begin{equation}
\ddot{\bfm R}+\bfm g(\bfm R)+\bfm f(\bfm R)\cos{t} =0\,
\label{eq::EOMdl}
\end{equation}
with  $\bfm g=\bfm G/(Lm\omega^2)$ and $\bfm f=\bfm
F/(Lm\omega^2)$. Note that in the rescaled variables $c=1$ and
the expansion parameter is  $v$. While $\bfm  g=O(v^2)$ by
definition, the scaling of the oscillating term needs to be
determined. The leading contribution of the oscillating field
to the effective action is quadratic in its amplitude and we
are interested in the physical systems where the large-scale
dynamics is essentially determined  by the effect of the
periodic drive, which should be comparable to the one of the
static field.  This requires $\bfm  f=O(v)$, {\it i.e.} with
the rest of the parameters fixed, the amplitude $\bfm F$ of the
oscillating field should  scale linearly with its frequency.
This does not necessarily mean the actual dependence of the
amplitude on the frequency but rather determines the relevant
range for the  oscillating field amplitude at a given $\omega$.
This corresponds to the ``strong-drive'' scaling relevant for
the study of the dynamical stabilization. The ``fixed-drive''
regime, where the amplitude $\bfm F$ remains constant in the
high-frequency limit, implies $\bfm  f=O(v^2)$ and can be
obtained from the strong-drive result by dropping the higher
powers of $\bfm  f$ field in the effective action at a given
order of the expansion.

A number of  methods  have been developed to disentangle the
slow and fast dynamics in perturbation theory (see
\cite{Krylov:1949,Bogoliubov:1961,Nayfeh:2008,Blekhman:2000,
Fidlin:2010,Thomsen:2021} for a review). They share the
principal idea of introducing independent variables for the
fast and slow evolution with subsequent averaging over the fast
one. Its particular realization, however, is crucial to get an
efficient tool for the high-order analysis. A natural choice of
the slow variable is the path along the smeared trajectory
\cite{Penin:2023dtn}. In this case the solution of
Eq.~(\ref{eq::EOMdl})  can be formally written as
\begin{eqnarray}
{\bfm R}&=&{\bfm r}+\delta {\bfm R}\,,
\nonumber\\
\delta{\bfm R}&=&\sum_{n=1}^\infty
\left[\bfm c_n({\bfm r})\cos(nt)
+\bfm s_n({\bfm r})\sin(nt)\right]\,,
\label{eq::modes}
\end{eqnarray}
where the vector ${\bfm r}(t)$ describes the smeared slow
motion. Then the total time derivative splits into the slow and
fast components as follows
\begin{equation}
{d\over dt}={\bfm
v}{\bfm\partial}+\partial_t\,,
\label{eq::dtsplit}
\end{equation}
where ${\bfm\partial}=\partial/\partial {\bfm
r}$.\footnote{Throughout the paper for every term with a pair
of the boldface $d$-dimensional vectors their scalar product
is implied.} Strictly  speaking our choice of the slow variable
is well defined only for ${\bfm v}\ne 0$.  The resulting
effective Lagrangian, however, is  regular in the limit of
vanishing velocity and can be continued to ${\bfm v}=0$. The
slow variation of the ${\bfm r}$ variable is manifested by the
requirement that its time derivative ${\bfm v}=\dot{\bfm r}$
scales as ${\cal O}(v)$ and is a function of ${\bfm r}$
only.\footnote{Here $v$ is the dimensionless ratio of the
parameters of the system defined above. The function ${\bfm
v}({\bfm r})$ describes the time dependence of the velocity on
a smeared trajectory ${\bfm r}(t)$ and does not impose any
constraint on the independent dynamical variables in the
effective theory.}  Thus the average acceleration ${\bfm
a}\equiv \ddot{\bfm r}=v_i\partial_i{\bfm v}$ is  ${\cal
O}(v^2)$.\footnote{The summation over the repeating indices is
implied throughout the paper and
$\partial_i\equiv\partial/\partial {r_i}$.} By substituting
Eqs.~(\ref{eq::modes},\,\ref{eq::dtsplit}) into
Eq.~(\ref{eq::EOMdl}), after the expansion in $v$ and in the
Fourier harmonics one gets the algebraic equations on the
harmonic coefficients ${\bfm c_n},~{\bfm s_n}$ and the
differential equation which determines the time evolution of
${\bfm r}$ order-by-order in the high-frequency expansion. The
resulting equation on ${\bfm r}$ is of the form ${\bfm a}+{\bfm
{\cal F}}_{\rm eff}=0$, and can be understood as the effective
theory equation of motion which governs the slow large-scale
evolution of the system under the action of the effective force
$-{\bfm {\cal F}}_{\rm eff}({\bfm r},{\bfm v})$.\footnote{The
negative sign is introduced for convenience and we will refer
to ${\bfm {\cal F}}_{\rm eff}$ as the effective force in the
rest of the paper.} The resulting function ${\bfm r}(t)$
determines the solution of Eq.~(\ref{eq::EOMdl}) through
Eq.~(\ref{eq::modes}). This approach is reminiscent of the
renowned  Krylov-Bogoliubov asymptotic method
\cite{Krylov:1949,Bogoliubov:1961} which has been originally
developed for the nonlinear oscillation theory. There are two
main distinction though: (i) the slow evolution is not limited
to the fast-oscillation amplitude variation, hence
Eq.~(\ref{eq::modes}) accommodates the zero harmonic, (ii) the
fast modes are determined by the external force, hence the
corrections to the fast-oscillation frequency are absent.

In practice the high-frequency expansion within this framework
is realized through the following steps. \\
{\bf 1.} We define the series
\begin{eqnarray}
{\bfm v}&=&\sum_{m=1}^\infty
{\bfm v}^{(m)}({\bfm r})\,,
\qquad
{\bfm a}=\sum_{m=2}^\infty
{\bfm a}^{(m)}({\bfm r})\,,
\nonumber\\
{\bfm c}_n&=&\sum_{m=1}^\infty {\bfm
c}_n^{(m)}({\bfm r})\,,\qquad \!
{\bfm s}_n=\sum_{m=1}^\infty {\bfm
s}_n^{(m)}({\bfm r})\,,
\label{eq::seriesv}
\end{eqnarray}
where the $m$th term of each series uniformly scales as ${\cal
O}(v^m)$. Note that the series coefficients of the acceleration
and velocity are related as follows
\begin{equation}
{\bfm a}^{(m)}=\sum_{l=1}^{m-1}
v_i^{(l)}\partial_i{\bfm v}^{(m-l)}\,,
\label{eq::seriesa}
\end{equation}
and the contribution of the higher harmonics is suppressed:
${\bfm c}_n^{(m)},{\bfm s}_n^{(m)}=0$ for $m<n$. \\ {\bf 2.}
Eqs.~(\ref{eq::modes},\,\ref{eq::dtsplit},\,\ref{eq::seriesv}) are
substituted into Eq.~(\ref{eq::EOMdl}) where the functions
${\bfm g}$ and ${\bfm f}$ are expanded in the series
\begin{eqnarray}
&&{\bfm g}({\bfm R})={\bfm g}({\bfm r})
+\sum_{n=1}^\infty {1\over n!}
\left(\delta R_{i_n}\ldots\delta R_{i_1}\right)
\partial_{i_n}\ldots\partial_{i_1}
{\bfm g}({\bfm r})\,,
\nonumber\\
&&{\bfm f}({\bfm R})={\bfm f}({\bfm r})
+\sum_{n=1}^\infty {1\over n!}
\left(\delta R_{i_n}\ldots\delta R_{i_1}\right)
\partial_{i_n}\ldots\partial_{i_1}
{\bfm f}({\bfm r})\,.
\nonumber\\[-2mm]
&&\label{eq::seriesgf}
\end{eqnarray}
Then, Eq.~(\ref{eq::seriesa}) is used to combine the velocity
gradient terms into the acceleration at a given order of the
expansion. \\
{\bf 3.} After the rescaling ${\bfm g}\to \varepsilon^2{\bfm
g}$, ${\bfm f}\to\varepsilon{\bfm f}$, $\{{\bfm v}^{(m)},{\bfm
a}^{(m)},{\bfm c}_n^{(m)}, {\bfm s}_n^{(m)}\}\to
\varepsilon^m\{{\bfm v}^{(m)},{\bfm a}^{(m)},{\bfm c}_n^{(m)},
{\bfm s}_n^{(m)}\}$, the resulting Eq.~(\ref{eq::EOMdl}) is
expanded  in the power series in  $\varepsilon$ and in Fourier
harmonics. Equating  the contribution of each Fourier harmonic
to zero in  $m$th order of $\varepsilon$-expansion provides a set
of equations for ${\bfm c}_n^{(m)}, {\bfm s}_n^{(m)}$ and
${\bfm a}^{(m)}$ in terms of the lower order coefficients. The
equations on the harmonic coefficients  ${\bfm c}_n^{(m)},
{\bfm s}_n^{(m)}$ are {\it algebraic} since their leading
contribution is determined by the fast $\partial_t$ part of the
time derivative, Eq.~(\ref{eq::dtsplit}), acting on $\cos(nt)$ or
$\sin(nt)$ functions in Eq.~(\ref{eq::modes}). The remaining
equation ${\bfm a}^{(m)}+{\bfm {\cal F}}_{\rm eff}^{(m)}=0$
defines  the effective force at ${\cal O}(v^m)$.\\
{\bf 4.} The $m$th-order equations may include the functions
${\bfm a}^{(i)}$ with $i<m$, which are eliminated by the lower
order equations of motion. Moreover, the equations in general
include the higher-order time derivatives $\dot{\bfm
a},~\ddot{\bfm a},\ldots$ which in turn are  eliminated
through the relations $\dot{\bfm a}=v_i\partial_i{\bfm {\cal
F}}_{\rm eff}$ {\it etc.} so that the effective force is always
a function of ${\bfm r}$ and ${\bfm v}$ only.

Let us now present the  results of the explicit calculation
through the first few orders  of the high-frequency expansion.
At ${\cal O}(v)$ we get
\begin{equation}
{\bfm c}_1^{(1)}={\bfm f}\,,
\qquad {\bfm s}_1^{(1)}=0\,,
\qquad {\bfm {\cal F}}_{\rm eff}^{(1)}=0\,.
\label{eq::v1}
\end{equation}
At  ${\cal O}(v^2)$ the harmonic coefficients read
\begin{equation}
\begin{array}{ll}
{\bfm c}_1^{(2)}=0\,, \quad&
{\bfm s}_1^{(2)}=-2v^{(1)}_i\partial_i {\bfm f}\,,\\[2mm]
{\bfm c}_2^{(2)}={1\over 8}f_i\partial_i {\bfm f}\,,
\quad&{\bfm s}_2^{(2)}=0\,,
\end{array}
\label{eq::v2}
\end{equation}
and the effective force is given by
\begin{equation}
\begin{split}
& {\bfm {\cal F}}_{\rm eff}^{(2)}={\bfm g}
+{1\over 2}f_i\partial_i {\bfm f}\,.
\label{eq::Feffv2}
\end{split}
\end{equation}
Note that Eq.~(\ref{eq::Feffv2}) is valid for  arbitrary
functions ${\bfm g}$ and ${\bfm f}$. If the corresponding
potentials exist so that $\bfm g={\bfm \partial}V_g$ and $\bfm
f={\bfm \partial}V_f$, then Eq.~(\ref{eq::Feffv2}) can be
understood as the Euler-Lagrange equation for the effective
Lagrangian
\begin{equation}
\begin{split}
&{\cal L}^{(2)}_{\rm eff}= {{\bfm v}^2\over 2}
-V^{(2)}_{\rm eff}\,, \qquad
V^{(2)}_{\rm eff}=V_g+{{\bfm f}^2\over 4}\,.
\label{eq::Leffv2}
\end{split}
\end{equation}
At {${\cal O}(v^3)$} the non-zero harmonic coefficients are
\begin{eqnarray}
{\bfm c}_1^{(3)}&=&3g_i\partial_i{\bfm f}
+f_i\partial_i{\bfm g}
+{3\over 8}f_if_j\partial_i\partial_j{\bfm f}
-3v^{(1)}_iv^{(1)}_j\partial_i\partial_j{\bfm f}
\nonumber\\
&+&{25\over 16}f_i(\partial_if_j)\partial_j{\bfm f}\,,
\nonumber\\
{\bfm s}_1^{(3)}&=&-2v^{(2)}_i\partial_i {\bfm f}\,,
\nonumber\\
{\bfm s}_2^{(3)}&=&-{1\over 8}v^{(1)}_if_j
\partial_i\partial_j {\bfm f}
-{3\over 8}v^{(1)}_i\left(\partial_if_j\right)
\partial_j {\bfm f}\,,
\nonumber\\
{\bfm c}_3^{(3)}&=&{1\over 72}f_if_j
\partial_i\partial_j {\bfm f}
+{1\over 144}f_i\left(\partial_if_j\right)
\partial_j {\bfm f}\,,
\label{eq::v3}
\end{eqnarray}
while the effective force vanishes ${\bfm {\cal F}}_{\rm
eff}^{(3)}=0$. The expressions for the ${\cal O}(v^4)$ harmonic
coefficients become rather lengthy and are given in the
Appendix~\ref{sec::appA}. Starting with  this order the
effective force  is velocity-dependent
\begin{eqnarray}
&&{\bfm {\cal F}}^{(4)}_{\rm eff}=-{3\over 2}v^{(1)}_iv^{(1)}_j
\left(\partial_i\partial_jf_k\right)\partial_k{\bfm f}
+{1\over 4}f_if_j\partial_i\partial_j {\bfm g}
\nonumber\\
&&+\bigg[{3\over 2}g_i\left(\partial_if_k\right)
+{1\over 2}f_i(\partial_ig_k)
+{25\over 32}f_i(\partial_if_j)(\partial_jf_k)
\nonumber\\
&&+{3\over 16}f_if_j(\partial_i\partial_jf_k)\bigg]
\partial_k{\bfm f}
+{1\over 32}f_if_j(\partial_if_k)
\partial_j\partial_k{\bfm f}
\nonumber\\
&&+{1\over 16}f_if_jf_k
\partial_i\partial_j\partial_k{\bfm f}\,,
\label{eq::Feffv4}
\end{eqnarray}
and the corresponding  Lagrangian reads
\begin{equation}
{\cal L}_{\rm eff}^{(4)}=
-{3\over 4}v_iv_j(\partial_i{\bfm f})\partial_j{\bfm f}
-V^{(4)}_{\rm eff}\,,
\label{eq::Leffv4}
\end{equation}
with the effective potential
\begin{equation}
\begin{split}
&{V}_{\rm eff}^{(4)}={1\over 4}f_i{\bfm f}\partial_i{\bfm g}
+{1\over 64}f_if_j(\partial_i{\bfm f})\partial_j{\bfm f}
+{1\over 16}f_if_j{\bfm f}\partial_i\partial_j{\bfm f}\,.
\label{eq::Veffv4}
\end{split}
\end{equation}
The higher-order harmonic coefficients become too cumbersome
and we present only the physically relevant results for the
effective force.  At ${\cal O}(v^5)$ it reads
\begin{equation}
{\bfm {\cal F}}_{\rm eff}^{(5)}=-{3}v^{(1)}_iv^{(2)}_j
\left(\partial_i\partial_jf_k\right)\partial_k{\bfm f}\,,
\label{eq::Feffv5}
\end{equation}
and can be directly obtained from
Eqs.~(\ref{eq::Feffv4},\,\ref{eq::Leffv4}) by keeping the
higher-order terms in the expansion of the velocity. The ${\cal
O}(v^6)$ expression for the force is given in the
Appendix~\ref{sec::appB}. The construction of the corresponding
effective  Lagrangian, however, is more subtle and is described
in the Appendix~\ref{sec::appC}. The  equation of motion with
the complex dependence of the effective force on the velocity
can only be transformed into the Euler-Lagrange form by a
velocity-dependent coordinate transformation ${\bfm r}\to {\bfm
r}'({\bfm r},{\bfm v})$ with
\begin{equation}
{\bfm r}'={\bfm r}+{5\over 3}v^{(1)}_iv^{(1)}_j
\left[(\partial_i\partial_jf_k){\bfm \partial}f_k
-(\partial_i{\bfm \partial}f_k)\partial_jf_k\right]\,.
\label{eq::rp}
\end{equation}
Note that since the correction term vanishes at zero velocity,
the difference between  ${\bfm r}$ and ${\bfm r}'$ vanishes for
the system at an equilibrium.  Adding up all the
contributions through ${\cal O}(v^6)$  in new variables we
obtain the effective Lagrangian in the next-to-next-to-leading
order of the high-frequency expansion
\begin{eqnarray}
&&{\cal L}_{\rm eff}={v'_iv'_j\over 2}\bigg\{\delta_{ij}
-{3\over 2}(\partial_i{\bfm f})\partial_j{\bfm f}
+{5\over 6}v'_kv'_l\left[2(\partial_i\partial_j\partial_k{\bfm f})
\partial_l{\bfm f}\right.
\nonumber\\
&&\left.-(\partial_i\partial_j{\bfm f})
\partial_k\partial_l{\bfm f}\right]
+{5\over 3}\left(2g_k+f_l\partial_lf_k\right)
\left[(\partial_i\partial_j{\bfm f})\partial_k{\bfm f}
\right.
\nonumber\\[2mm]
&&\left.
-(\partial_i\partial_k{\bfm f})\partial_j{\bfm f}\right]
-5g_k(\partial_k{\bfm f})\partial_i\partial_j{\bfm f}
-5g_k(\partial_i{\bfm f})\partial_j\partial_k{\bfm f}
\nonumber\\[2.5mm]
&&-5(\partial_if_k)(\partial_k{\bfm g})\partial_j{\bfm f}
-5(\partial_if_k)(\partial_j{\bfm g})\partial_k{\bfm f}
-3f_k(\partial_j{\bfm f})\partial_i\partial_k{\bfm g}
\nonumber\\[1mm]
&&-{5\over 2}f_k(\partial_kf_l)(\partial_l{\bfm f})
\partial_i\partial_j{\bfm f}
-{43\over 16}f_k(\partial_kf_l)(\partial_i{\bfm f})
\partial_l\partial_j{\bfm f}
\nonumber\\
&&-{11\over 4}f_k(\partial_if_l)
(\partial_j{\bfm f})\partial_k\partial_l{\bfm f}
-{183\over 64}f_k(\partial_if_l)
(\partial_l{\bfm f})\partial_j\partial_k{\bfm f}
\nonumber\\
&&-{3\over 128}f_kf_l(\partial_i\partial_k{\bfm f})
\partial_j\partial_l{\bfm f}
-{9\over 8}f_kf_l(\partial_i{\bfm f})
\partial_j\partial_k\partial_l{\bfm f}
\nonumber\\
&&-{379\over 128}(\partial_if_k)(\partial_kf_l)
(\partial_l{\bfm f})\partial_j{\bfm f}
\bigg\}
\nonumber\\
&&-V_{\rm eff}+{\cal O}(v^8)\,,
\label{eq::Leffnnlo}
\end{eqnarray}
where   ${\bfm v}'=\dot{\bfm r}'$,
${\bfm\partial}=\partial/\partial {\bfm r}'$, and the effective
potential reads
\begin{eqnarray}
&&V_{\rm eff}=V_g
+{{\bfm f}^2\over 4}
+{1\over 4}f_i{\bfm f}\partial_i{\bfm g}
+{1\over 64}f_if_j(\partial_i{\bfm f})\partial_j{\bfm f}
\nonumber\\
&&
+{1\over 16}f_if_j{\bfm f}\partial_i\partial_j{\bfm f}
-{5\over 4}g_ig_j(\partial_i{\bfm f})\partial_j{\bfm f}
\nonumber\\
&&
+{1\over 4}f_if_j(\partial_i{\bfm g})\partial_j{\bfm g}
-{5\over 4}g_if_j(\partial_i{\bfm f})
(\partial_jf_k)\partial_k{\bfm f}
\nonumber\\
&&
+{1\over 256}f_if_j(\partial_if_k)
(\partial_j{\bfm f})\partial_k{\bfm g}
+{1\over 32}f_if_j(\partial_if_k)
(\partial_k{\bfm f})\partial_j{\bfm g}
\nonumber\\
&&
+{3\over 16}f_if_jf_k(\partial_i\partial_j{\bfm f})
\partial_k{\bfm g}
+{1\over 64}f_if_jf_k{\bfm f}
\partial_i\partial_j\partial_k{\bfm g}
\nonumber\\
&&
+{1\over 64}f_if_jf_k(\partial_i{\bfm f})
\partial_j\partial_k{\bfm g}
-{1435\over 4608}f_if_j(\partial_i{\bfm f})
(\partial_jf_k)(\partial_kf_l)\partial_l{\bfm f}
\nonumber\\
&&
+{1\over 512}f_if_jf_k(\partial_i{\bfm f})
(\partial_jf_l)\partial_k\partial_l{\bfm f}
+{7\over 576}f_if_jf_k(\partial_jf_l)
(\partial_l{\bfm f})
\nonumber\\
&&\times\partial_i\partial_k{\bfm f}
+{41\over 1152}f_if_jf_kf_l
(\partial_i\partial_j{\bfm f})
\partial_k\partial_l{\bfm f}
+{1\over 192}f_if_jf_kf_l
\nonumber\\
&&\times(\partial_i{\bfm f})\partial_j
\partial_k\partial_l{\bfm f}
+{1\over 384}f_if_jf_kf_l{\bfm f}
\partial_i\partial_j\partial_k\partial_l{\bfm f}
\,.
\label{eq::Veffnnlo}
\end{eqnarray}

A few comments  are in order.  First, due to the time-reversal
invariance of Eq.~(\ref{eq::EOMdl}) the high-frequency
expansion is in even powers of $1/\omega$. Then, to facilitate
the practical applications  it is instructive to transform the
result into the physical  units. Through the next-to-leading
order the effective Lagrangian becomes
\begin{eqnarray}
&&{\cal L}_{\rm eff}=m{v_iv_j\over 2}\bigg(\delta_{ij}
-{3\over 2}{(\partial_i{\bfm F})\partial_j{\bfm F}
\over m^2 \omega^4}\bigg)
-\bigg(V_g +{1\over 4}{{\bfm F}^2\over m \omega^2}
\nonumber\\
&&+{1\over 4}{F_i{\bfm F}\partial_i{\bfm G}\over m^2\omega^4}
+{1\over 64}{F_iF_j(\partial_i{\bfm F})\partial_j{\bfm F}
\over m^3\omega^6}
+{1\over 16}{F_iF_j{\bfm F}\partial_i\partial_j{\bfm F}
\over m^3\omega^6} \bigg).
\nonumber\\
&&
\label{eq::phunits}
\end{eqnarray}
In higher orders  the physical units can be restored by
introducing an extra $1/(m\omega^2)$ factor per each additional
power of the spatial derivative, and an extra $m\omega$ factor
per each additional power of velocity, so that
$v_i\partial_i\to v_i\partial_i/\omega$. Thus, assuming smooth
forces,  the expansion parameter for the effective potential in
the physical units is given by a dimensionless ratio of the
${\bfm F}$ and ${\bfm G}$ field  gradients to $m\omega^2$. Note
that our power counting corresponds to the strong-drive regime
with ${\bfm F}={\cal O}(\omega)$. In the fixed-drive regime
with ${\bfm F}={\cal O}(1)$ the last two terms in
Eq.~(\ref{eq::phunits}) become next-to-next-to-leading and
should be omitted.

Finally, we would like to note that the above ${\cal O}(v^6)$
result may be important for the high-precision analysis of the
parametric resonances in the rapidly driven systems
\cite{Chirikov:1979} and, in particular, of the instabilities
due to the overlapping resonances
\cite{Ugulava:2003,Ugulava:2004,Wang:2018,Plestid:2019,
Sarkar:2021,Luo:2022,Wanic:2024,Deng:2025}. This, however, is
beyond the scope of the present work.

\vspace*{3mm}
\section{Dynamical stabilization in one dimension}
\label{sec::2}

\vspace*{2mm}

The structure of the effective theory greatly simplifies in one
dimensional systems. In particular the correction term in
Eq.~(\ref{eq::rp}) vanishes and no coordinate transformation is
necessary in  the derivation of the effective Lagrangian. For
a generalized coordinate $q$ and arbitrary functions $g$ and $f$
it takes the following form through ${\cal O}(v^6)$
\begin{eqnarray}
&&{\cal L}_{\rm eff}={\dot{q}^2\over 2}
\bigg[1-{3\over 2}f'^2
+{5\over 6}\dot{q}^2\left(2f'f'''-f''^2\right)
-10f'f''g
\nonumber\\
&&-10f'^2g'-3ff'g''-{379\over 128}f'^4
-{691\over 64}ff'^2f''-{3\over 128} f^2f''^2
\nonumber\\
&&
-{9\over 8}f^2f'f'''\bigg]-V_{\rm eff}+{\cal O}(v^8)\,,
\label{eq::Leff1dnnlo}
\end{eqnarray}
where dash stands for the derivative $d/dq$ and
the effective potential reads
\begin{eqnarray}
&&V_{\rm eff}=V_g+{f^2\over 4}+{f^2g'\over 4}
+{f^2f'^2\over 64}+{f^3f''\over 16}
\nonumber\\
&&-{5\over 4}f'^2g^2+{f^2g'^2\over 4}
-{5\over 4}ff'^3g+{9\over 256}f^2f'^2g'
+{3\over 16}f^3f''g'
\nonumber\\
&&+{f^3f'g''\over 64}+{f^4g'''\over 64}
-{1435\over 4608}f^2f'^4
+{65\over 4608}f^3f'^2f''
\nonumber\\
&&+{41\over 1152}f^4f''^2
+{f^4f'f'''\over 192}+{f^5f''''\over 384}
\,.
\label{eq::Veff1dnnlo}
\end{eqnarray}
Eqs.~(\ref{eq::Leff1dnnlo},\,\ref{eq::Veff1dnnlo}) can be
obtained from the $d$-dimensional expressions
Eqs.~(\ref{eq::Leffnnlo},\,\ref{eq::Veffnnlo}) by dimensional
reduction and agree with \cite{Penin:2023dtn}.

The above result is of a particular
interest since many physical systems can be reduced to or
decomposed  into the one-dimensional problems. The high-order
result is not only mandatory for  establishing the convergence
of the high-frequency expansion and increasing the  accuracy of
the theoretical predictions, but  also reveals nontrivial
general features of the dynamical stabilization of nonlinear
systems. Let us discuss one of them. The dynamical
stabilization is a phenomenon when the presence of the rapidly
oscillating field changes the stability of the static system.
Let us assume that the static system is  in equilibrium at
$q=0$. Excluding the degenerate cases this implies
$g(q)=\delta q+{\cal O}(q^2)$ with a constant $\delta$. The
rapidly oscillating force generates the effective force which
may change the stability of the system when
\begin{equation}
\partial _q{{\cal F}}_{\rm eff}\big|_{q=\dot{q}=0} =0\,.
\label{eq::bif}
\end{equation}
If  $f$ depends on a parameter $\epsilon$, Eq.~(\ref{eq::bif})
defines $\delta$ as a function of $\epsilon$, {\it i.e.} a
transition curve in the parameter space of the system. Crossing
such a curve results in a bifurcation and qualitative change in
the system behavior. For $g(q)=\delta q$, $f(q)=\epsilon  q$
the problem is reduced to the  linear Mathieu equation, where
the  transition curves can be obtained {\it e.g.} by the method
of harmonic balance (see \cite{Kovacic:2018} for a review) or
homotopy analysis \cite{Desai:2023}  beyond the perturbation
theory for arbitrary finite values of  $\delta$ and $\epsilon$.
In general, for the nonlinear systems the corresponding
transition curves are obviously different. Moreover, the
high-order contributions to the effective potential depend on
the high-order derivatives of the functions $g$ and $f$, which
makes it more and more sensitive to the nonlinear behavior of
the  forces near the equilibrium position.

There is, however, one remarkable exception.  Let us consider
an equilibrium point of the static system at $q=0$, so that
$g(q)=\delta q+{\cal O}(q^2)$. By construction, for all the
terms in the effective potential the total number of the
derivatives is less than the total power of the $g$ and $f$
functions by two, {\it cf.} Eq.~(\ref{eq::Veff1dnnlo}).   Thus
one immediately finds that  if  $f(0)=0$, {\it i.e.}
$f(q)=\epsilon q+{\cal O}(q^2)$, then the contribution of the
nonlinear terms in the expansion of $g(q)$ and $f(q)$ about
$q=0$  to Eq.~(\ref{eq::bif}) vanishes and the corresponding
transition curves coincide with those of the linear  Mathieu
equation to all orders in the high-frequency expansion. The
left hand side of Eq.~(\ref{eq::bif}) for the linear Mathieu
equation is known to be an analytic function of $\delta$ and
$\epsilon$ near the origin $\delta=\epsilon=0$
\cite{Arscott:1964}, and the corresponding Taylor series has
a finite radius of convergence (the explicit result for the
series up to ${\cal O}(\epsilon^{28})$ as
well as the convergence radius estimate can be found  in
\cite{Beneke:2023ndu}). Hence, by analytic continuation the
equivalence of the transition curves can be extended beyond the
high-frequency expansion to arbitrary values of $\delta$ and
$\epsilon$. This proves the following nontrivial  theorem:

\vspace*{2.5mm}
\noindent
\fbox{\parbox[c]{0.46\textwidth}{\em   For an arbitrary
nonlinear system, if the amplitude of the rapidly oscillating
force vanishes at an equilibrium point of the static
interaction, the corresponding transition curves coincide with
those of the linear  Mathieu equation.}}

\vspace*{2.5mm}
\noindent
We should emphasize that if the above condition is not
satisfied, Eq.~(\ref{eq::bif}) for a nonlinear system differs
from the one of its linearized approximation near the static
equilibrium point starting from the next-to-leading order, {\it
i.e.} the linearization and time-averaging {\it do not
commute}.  An interesting problem in this context is to study
how a small deviation of $f(0)$ from zero perturbs the
transition curves for nonlinear systems, which, however, is
beyond the scope of this paper.

The equivalence of the transition curves extends all the
nontrivial features of the Mathieu equation stability chart
\cite{Kovacic:2018,Ince:1927} to the above class of the
nonlinear systems with arbitrary  nonlinearity and  driving
force amplitude. The class of the system  satisfying the above
condition is quite broad. In particular, it includes all the
one-dimensional systems with vanishing static force $g=0$ for
dynamical equilibrium at the zeros of the amplitude $f$. The
generalization of the  analysis to multiple spatial dimensions
is more subtle. Indeed, the stability chart of an
$d$-dimensional non-linear system satisfying the above
condition reduces  to that of a linear system in the same way.
However, already the two-dimensional systems may have a more
complex stability chart than the Mathieu equation and should be
studied case-wise (see the example at the end of
Sect.~\ref{sec::4}).

A very illustrative application of this theorem is the Kapitza
pendulum. For the vertically oscillating pivot we  have
$g=\delta \sin(\theta)$, $f=\epsilon \sin(\theta)$, where the
dimensionless parameters are $\delta={\rm g}/(L\omega^2)$,
$\epsilon =(\Delta/L)$. Here $\theta$ is the angle measured
from the downward position, $L$ is the pendulum length, ${\rm
g}$ is the free fall acceleration, $\Delta$ and $\omega$ are
the pivot oscillation amplitude and frequency, respectively.
Note that at the static unstable equilibrium point $\theta=\pi$
the oscillation amplitude vanishes, so that the system belongs
to the class discussed above. For the effective potential we
get
\begin{eqnarray}
V^{(2)}_{\rm eff}&=&-\delta\cos(\theta) -
{\epsilon^2\over 8}\cos(2\theta)\,,
\nonumber\\
V^{(4)}_{\rm eff}&=&
{1\over 16}\delta\epsilon^2\cos(\theta)
+{1\over 32}\epsilon^4\cos(2\theta)
\nonumber\\
&-&{1\over 16}\delta\epsilon^2\cos(3\theta)
-{5\over 512}\epsilon^4\cos(4\theta)\,,
\nonumber\\
V^{(6)}_{\rm eff}&=&
-{367\over 2048} \delta\epsilon^4\cos(\theta)
-{1993\over 73728}\epsilon^6\cos(2\theta)
\nonumber\\
&+&{479\over 4096}\delta\epsilon^4\cos(3\theta)
+\left({1\over 8}\delta^2\epsilon^2
+{57\over 2048}\epsilon^6\right)\cos(4\theta)
\nonumber\\
&+&{255\over 4096}\delta\epsilon^4\cos(5\theta)
+{65\over 8192}\epsilon^6\cos(6\theta)\,.
\label{eq::Veffnnlovp}
\end{eqnarray}
Substituting it into Eq.~(\ref{eq::bif}) we get
\begin{equation}
\begin{split}
& \delta={\epsilon^2\over 2}-{7\over 32}\epsilon^4
+{29\over 144}\epsilon^6+{\cal O}(\epsilon^8)\,,
\label{eq::stabilityvp}
\end{split}
\end{equation}
which coincides with the first transition curve of the  Mathieu
equation \cite{Kovacic:2018}. It describes a pitchfork
bifurcation corresponding to the dynamical stabilization of the
upward equilibrium position discussed in the original work
\cite{Kapitza:1951}.  The  corrections to the
effective potential have been analysed using different methods
in \cite{Venkatraman:2021lwv,Beneke:2023ndu}. Upon the use of
the equation of motion,
Eqs.~(\ref{eq::Leff1dnnlo},\,\ref{eq::Veffnnlovp}) agree with
the next-to-leading effective Lagrangian presented in
\cite{Beneke:2023ndu}. In higher orders the comparison of the
results is not straightforward since in \cite{Beneke:2023ndu}
the velocity-dependent terms are eliminated from  the equation
of motion by using the energy conservation. Hence, the
resulting  effective potential depends on  the total energy of
the system, while we use the standard definition of the
Lagrangian  independent of the initial conditions. At the same
time, the transition curve derived from the result
\cite{Beneke:2023ndu} agrees with Eq.~(\ref{eq::stabilityvp})
through the next-to-next-to-leading order.  The method of
\cite{Venkatraman:2021lwv} relies on a canonical transformation
to decouple the fast modes and  construct the  effective
static Hamiltonian for a rapidly driven system. Our result
disagrees with the next-to-leading order Hamiltonian for the
Kapitza pendulum presented in this work.

By contrast, for the horizontally oscillating pivot we have
$f=\epsilon \cos(\theta)$  so that for the static stable
equilibrium point $\theta=0$ the fast oscillation amplitude
does not vanish. In this case the effective potential reads
\begin{eqnarray}
V^{(2)}_{\rm eff}&=&-\delta\cos(\theta) +
{\epsilon^2\over 8}\cos(2\theta)\,,
\nonumber\\
V^{(4)}_{\rm eff}&=&
{3\over 16}\delta\epsilon^2\cos(\theta)
-{1\over 32}\epsilon^4\cos(2\theta)
\nonumber\\
&+&{1\over 16}\delta\epsilon^2\cos(3\theta)
-{5\over 512}\epsilon^4\cos(4\theta)\,,
\nonumber\\
V^{(6)}_{\rm eff}&=&
{73\over 2048} \delta\epsilon^4\cos(\theta)
+\left({3\over 4}\delta^2\epsilon^2
+{1993\over 73728}\epsilon^6\right)\cos(2\theta)
\nonumber\\
&-&{1233\over 4096}\delta\epsilon^4\cos(3\theta)
+\left(-{1\over 8}\delta^2\epsilon^2
+{57\over 2048}\epsilon^6\right)\cos(4\theta)
\nonumber\\
&+&{255\over 4096}\delta\epsilon^4\cos(5\theta)
-{65\over 8192}\epsilon^6\cos(6\theta)
\,,
\label{eq::Veffnnlohp}
\end{eqnarray}
and defines a different transition curve
\begin{equation}
\begin{split}
&\delta={\epsilon^2\over 2}+{3\over 32}\epsilon^4
+{17\over 576}\epsilon^6+{\cal O}(\epsilon^8)\,,
\label{eq::stabilityhp}
\end{split}
\end{equation}
which  describes  a pitchfork bifurcation corresponding to the
dynamical destabilization of the downward equilibrium with the
appearance of two symmetric stable equilibria  at a nonzero
value of the polar angle.

Despite the similarity, the two above cases are principally
different for the stability analysis. For the classical setup
with the vertically oscillating pivot from the equivalence of
the transition curves we conclude that, as for the Mathieu
equation, for any $\delta$ there are infinitely many  narrow
stability islands of rapidly decreasing width for the inverted
position of the pendulum with infinitely growing $\epsilon$,
{\it i.e.} the amplitude of the pivot oscillation. The first
few of these islands or ``tongues'' have been observed by a
dedicated numerical study in \cite{Broer:2004} and indeed agree
with the Mathieu equation stability chart. On the contrary, for
the horizontally oscillating pivot (in general, for any
oscillation direction except strictly vertical) we cannot make
any conclusions about the dynamical stabilization beyond the
perturbation theory which implies $\epsilon\ll 1$. At the same
time we can use the result Eq.~(\ref{eq::Veffnnlohp}) to study
the actual convergence region of the high-frequency expansion.
The effective potential at the near-critical value
$\delta={\epsilon^2/ 2}$ in different approximations is plotted
in Fig.~\ref{fig::1}, which clearly indicates breakdown of the
perturbation theory for $\epsilon\gsim 1/2$. The accuracy of
the approximation and the effect of the corrections for the
strongly nonlinear oscillations about the dynamically stable
equilibrium position close to $\theta=\pi/2$ for different
values of the expansion parameter are demonstrated in
Fig.~\ref{fig::2}.

\section{Configuration space with curvature}
\label{sec::3}
In Sect.~\ref{sec::1} the particle position in the
$d$-dimensional space was described by the Cartesian
coordinates ${\bfm R}$. In many applications, however, the
presence of the constraints requires introduction of the
generalized coordinates. Let us consider a system with the
contravariant generalized coordinates $Q^\mu$ in configuration
space of arbitrary dimension.\footnote{We use the Greek letter
indices in this section to distinguish the generalized
coordinates from the Cartesian coordinates of
Sect.~\ref{sec::1}.} Then the kinetic  energy can be written as
follows $g_{\mu\nu}(Q)\dot{Q}^\mu\dot{Q}^\nu/2$, where
$g_{\mu\nu}$ is a symmetric positively defined tensor which can
be understood as the metric tensor on the configuration space.
If $g_{\mu\nu}$ cannot be globally reduced to the Euclidean
form $\delta_{\mu\nu}$ by a coordinate transformation, the
configuration space has non-zero curvature and  the effective
theory analysis is to be modified.  The equations of motion in
this case take the form ({\it cf.} Eq.~(\ref{eq::EOMdl}))
\begin{equation}
\ddot{Q}^\mu+\Gamma^{\mu}_{\nu\lambda}(Q)\dot{Q}^\nu
\dot{Q}^\lambda+g^\mu(Q)+f^\mu(Q)\cos{\omega t}=0\,,
\label{eq::EOMfullq}
\end{equation}
where
\begin{equation}
\Gamma^{\mu}_{\nu\lambda}(Q)={g^{\mu\kappa}\over 2}
\left(\partial_\nu g_{\kappa\lambda}
+\partial_\lambda g_{\nu\kappa}
-\partial_\kappa g_{\nu\lambda} \right) \,,
\label{eq::Christ}
\end{equation}
are the Christoffel symbols,  $g^{\mu\nu}=g^{-1}_{\mu\nu}$,
$\partial_\mu=\partial/\partial Q^\mu$, and the contravariant
components of a force are obtained from the corresponding
potentials in the standard way, {\it e.g.}
$f^\mu=g^{\mu\nu}f_\nu$ with $f_\mu=\partial_\mu V_f$.

The  high-frequency expansion can be  performed
through the steps described in Sect.~\ref{sec::1}.
We substitute the decomposition
\begin{equation}
{Q^\mu}={q^\mu}+\sum_{n=1}^\infty
\left[c_n^\mu({q})\cos(nt)+s_n^\mu({q})\sin(nt)\right]\,,
\label{eq::modesq}
\end{equation}
and $d/d t=\dot{q}^\mu\partial_\mu +\partial_t$ into
Eq.~(\ref{eq::EOMfullq}) and get the equations for the harmonic
coefficients and the time evolution of the slow variables
$q_\mu$ order by order in inverse frequency. The main
difference with the previous analysis is in the presence of the
additional metric connection term. At ${\cal O}(v)$ we get
${c_1^\mu}^{(1)}=f^\mu$, and the ${\cal O}(v^2)$ effective
theory equation of motion becomes
\begin{equation}
\ddot{q}^\mu+\Gamma^{\mu}_{\nu\lambda}(q)\dot{q}^\nu
\dot{q}^\lambda+g^\mu(q)+{1\over 2}f^\nu(q) D_{\nu}f^\mu(q)=0\,,
\label{eq::EOMeffq}
\end{equation}
where $D_{\nu}f^\mu=\partial_{\nu}f^\mu
+\Gamma^{\mu}_{\nu\lambda}f^\lambda$ is the covariant
derivative. The corresponding effective Lagrangian reads
\begin{equation}
\begin{split}
&L_{\rm eff}={g_{\mu\nu}\over 2}\dot{q}^\mu\dot{q}^\nu
-V_{\rm eff}\,, \qquad V_{\rm eff}=V_g+{1\over 4}f^\mu f_\mu.
\label{eq::Leffq}
\end{split}
\end{equation}
The extension of the analysis to  higher orders is
straightforward.

As an example let us apply this covariant formalism to two
simplest   mechanical systems with the curved configuration
space. First, we consider a spherical pendulum with the pivot
oscillating  along the horizontal $x$-axis, where the
coordinates are the spherical angles $Q^\mu=(\theta,\phi)$, and
the metric tensor is
\begin{equation}
g_{\mu\nu}=
\left(\begin{array}{cc}
1 & 0 \\
0 &  \sin^2(\theta)
\end{array}
\right)\,.
\label{eq::metricsp}
\end{equation}
The exact equations of motion are
\begin{eqnarray}
&&\ddot{\theta}-\sin(\theta)\cos(\theta)\dot{\phi}^2
+\delta\sin(\theta)
+\epsilon\cos(t)\cos(\phi)\cos(\theta)=0,
\nonumber\\
&&\sin^2(\theta)\ddot{\phi}
+2\cos(\theta)\sin(\theta)\dot{\theta}\dot{\phi}
-\epsilon\cos(t){\sin(\phi)\sin(\theta)}=0,
\nonumber\\
&&\label{eq::EOMsp}
\end{eqnarray}
that defines the covariant components of the force
$f_\theta=\epsilon\cos(\phi)\cos(\theta)$, $f_\phi=-\epsilon
\sin(\phi)\sin(\theta)$.  The ${\cal O}(v^2)$ effective
potential then reads
\begin{equation}
V_{\rm eff}= -\delta \cos(\theta)
+{\epsilon^2\over 4}\left(\cos^2(\theta)\cos^2(\phi)
+\sin^2(\phi)\right)\,,
\label{eq::Veffsp}
\end{equation}
where we used $f^\mu f_\mu=g^{\theta\theta}f_\theta^2
+2g^{\theta\phi}f_\theta f_\phi+g^{\phi\phi}f_\phi^2$ for the
covariant square of the vector. The system undergoes a
pitchfork bifurcation at $\delta\approx{\epsilon^2/2}$ with the
appearance of two symmetric minima of the effective potential,
see Fig.~\ref{fig::3}a. The trajectories of the nonlinear
oscillations about the dynamical equilibrium position obtained
within the full and the effective theory are plotted in
Fig.~\ref{fig::4}a and Fig.~\ref{fig::4}b, respectively.

As the second example we consider a symmetric planar double
pendulum with the horizontally oscillating pivot.  In this case
the coordinates are the polar angles $\theta_1$ and $\theta_2$
of the upper and lower links, respectively, measured from the
vertical position. Thus, we have $Q^\mu=(\theta_1,\theta_2)$
and the corresponding  metric tensor can be easily computed
\begin{equation}
g_{\mu\nu}=
\left(\begin{array}{cc}
2 & \cos(\theta_1-\theta_2) \\
\cos(\theta_1-\theta_2) &  1
\end{array}
\right)\,.
\label{eq::metricdp}
\end{equation}
The exact equations of motion are
\begin{eqnarray}
&&2\ddot{\theta_1}+\cos(\theta_1-\theta_2)\ddot{\theta_2}
+\sin(\theta_1-\theta_2)\dot{\theta_2}^2+2\delta\sin(\theta_1)
\nonumber\\
&&+2\epsilon\cos(t)\cos(\theta_1)=0\,,
\nonumber\\
&&\ddot{\theta_2}+\cos(\theta_1-\theta_2)\ddot{\theta_1}
-\sin(\theta_1-\theta_2)\dot{\theta_1}^2+\delta\sin(\theta_2)
\nonumber\\
&&+\epsilon\cos(t)\cos(\theta_2)=0\,,
\label{eq::EOMdp}
\end{eqnarray}
where the parameters $\delta$ and $\epsilon$ are defined as
before, with $L$ and $m$ being the length and the mass of each
link of the pendulum. From Eq.~(\ref{eq::EOMdp}) we find the
covariant components
$f_{\theta_1}=2\epsilon\cos(\theta_1)$,
$f_{\theta_2}=\epsilon\cos(\theta_2)$, which along with
the metric Eq.~(\ref{eq::metricdp}) determine the ${\cal
O}(v^2)$ effective potential
\begin{eqnarray}
&&V_{\rm eff}= -\delta\left(2\cos(\theta_1)
+\cos(\theta_2)\right)+{\epsilon^2\over 2}
\nonumber\\
&&\times{2\cos^2(\theta_1)+\cos^2(\theta_2)
-2\cos(\theta_1-\theta_2)\cos(\theta_1)\cos(\theta_2)
\over 1+\sin^2(\theta_1-\theta_2)}\,.
\nonumber\\
&&\label{eq::Veffdp}
\end{eqnarray}
The stability chart of this system turns out to be quite
interesting. At  $\delta\approx{\epsilon^2/2}$ it undergoes a
pitchfork bifurcation similar to the one discussed for the
planar and spherical pendula with the dynamical equilibrium
configuration corresponding  to nonzero $\theta_1$ and
$\theta_2$  of the same sign. Then at
$\delta\approx{\epsilon^2/4}$ the second (this time fold)
bifurcation takes place and two extra symmetric local minima of
the effective potential appear, corresponding to   $\theta_1$
and $\theta_2$ of the opposite signs, see Fig.~\ref{fig::3}b.

\section{Dynamical magnetic confinement}
\label{sec::4}
In this section we extend the analysis to the systems
characterized by the presence of the velocity-dependent and
non-potential forces. The most physically interesting problem
in this class is the study of a charged particle motion in the
rapidly oscillating magnetic field and related problem of the
dynamical magnetic confinement \cite{Besharat:2025vbw}.

The equation of motion for  a particle of mass $m$ and electric
charge $e$ in the rapidly oscillating magnetic field ${\bfm
B}(t,{\bfm R})=\cos(\omega t){\bfm B}({\bfm R})$ and the
corresponding  induced  electric field  ${\bfm E}(t,{\bfm
R})=\sin(\omega t){\bfm E}({\bfm R})$, with  ${\bfm
\partial}\times{\bfm E}({\bfm R})=\omega{\bfm B}({\bfm R})$,
reads\footnote{In this section we keep the standard units to
facilitate the connection with the experiment.}
\begin{equation}
\ddot{\bfm R}-{e\over m}\left({\bfm E}(t,{\bfm R})
+\dot{\bfm R}\times {\bfm B}(t,{\bfm R})\right)=0\,.
\label{eq::EOMmag}
\end{equation}
The phase shift between the time-dependent forces, and their
non-potential character require an  adjustment of the analysis.
Eq.~(\ref{eq::EOMmag}) is invariant with respect to the time
reversal up to the overall phase shift. Hence, the
high-frequency expansion is in even powers of $1/\omega$ as in
the previous examples. Following the steps described in
Sect.~\ref{sec::1} through ${\cal O}(1/\omega^4)$ we obtain the
harmonic coefficients
\begin{eqnarray}
\bm{c}_1&=&-\frac{ev_j}{m \omega^3}\left(\bm{\d} E_j+\d_j \bm{E}\right)\,,
\nonumber\\
\bm{s}_1&=&-\frac{e\bm{E}}{m\omega^2}+\frac{e}{m\omega^4}
v_iv_j\left(\d_i\d_j\bm{E}+\bm{\d}\d_i E_j+\bm{\d}\d_j E_i\right)\,,
\nonumber\\
&&
\label{eq:4thEB}
\end{eqnarray}
and the effective force
\begin{eqnarray}
&&{\cal F}_{\rm eff}=
\frac{e^2}{2m^2\omega^2}E_i\bm{\d}E_i
+\frac{e^2}{2m^2\omega^4}v_jv_k\big(
-(\d_j\d_kE_i)\d_i\bm{E}
\nonumber\\[2mm]
&&-(\d_k\d_iE_j)\bm{\d}E_i
-(\d_k\d_iE_j)\d_i\bm{E}
+(\d_iE_j)\bm{\d}\d_i E_k
\nonumber\\[2mm]
&&
-(\d_iE_j)\d_k\d_i\bm{E}
+(\d_jE_i)\bm{\d}\d_iE_k
-(\d_jE_i)\d_k\d_i\bm{E}
\big)\,,
\nonumber\\
&&
\label{eq:EffEB}
\end{eqnarray}
where we kept the  terms  quadratic in the electric and
magnetic fields and eliminated the latter by the relation
${\bfm B} = {\bfm \partial}\times{\bfm E}/\omega$, so that the
result does not explicitly depend on ${\bfm B}$. The
corresponding effective Lagrangian reads
\begin{eqnarray}
{\cal L}_{\rm eff}&=&{m\over 2}v_iv_j\bigg[
\delta_{ij} -{e^2\over 2m^2\omega^4}
\left({\bfm \partial}E_i{\bfm \partial}E_j\right.
\nonumber\\
&+&\left.\partial_i{\bfm E}{\bfm \partial}E_j
+\partial_j{\bfm E}{\bfm \partial}E_i\right)
\bigg] -V_{\rm eff}\,,
\label{eq::LeffE}
\end{eqnarray}
with
\begin{equation}
V_{\rm eff}={e^2\over 4m\omega^2}{\bfm E}^2\,.
\label{eq::VeffE}
\end{equation}

As an important application of this result, let us consider
the  magnetic field of the commonly used Ioffe-Pritchard traps
\cite{Pritchard:1983}
\begin{equation}
{\bfm B}^{(z)}=B\left({xz\over\Delta^2}, {yz\over\Delta^2},
1+{x^2+y^2-2z^2\over 2\Delta^2}\right)\,,
\label{eq::Bfield}
\end{equation}
where  $B$ is the value of the homogeneous component of the
field (the superscript of ${\bfm B}^{(z)}$  indicates its
direction), and an adjustable parameter $\Delta$ defines  the
scale of the field variation determined by the trap geometry.
For the oscillating magnetic  field  with the amplitude
Eq.~(\ref{eq::Bfield}) the induced electric field amplitude
reads
\begin{equation}
{\bfm E}={\omega B\over 2}
\left(-y\left(1+{y^2-3z^2\over 3\Delta^2}\right),
x\left(1+{x^2-3z^2\over 3\Delta^2}\right),0\right)\,.
\label{eq::Efield}
\end{equation}
For this field configuration the exact equations of motion
are
\begin{eqnarray}
&&\ddot{x}=\omega_B\left\{-{\omega\, y\over 2}
\left(1+{y^2-3z^2\over 3\Delta^2}\right)\sin(\omega t)
\right.
\nonumber\\
&&\left.+\left[\dot{y}\left(1+{x^2+y^2-2z^2\over 2\Delta^2}\right)
-\dot{z}{yz\over \Delta^2}\right]\cos(\omega t)\right\}\,,
\nonumber\\
&&\ddot{y}=\omega_B\left\{{\omega\, x\over 2}
\left(1+{x^2-3z^2\over 3\Delta^2}\right)\sin(\omega t)
\right.
\nonumber\\
&&\left.+\left[-\dot{x}\left(1+{x^2+y^2-2z^2\over 2\Delta^2}\right)
+\dot{z}{xz\over \Delta^2}\right]\cos(\omega t)\right\}\,,
\nonumber\\
&&\ddot{z}=\omega_B\left(\dot{x}{yz\over \Delta^2}
-\dot{y}{xz\over \Delta^2}\right)\cos(\omega t)\,,
\label{eq::EOMtr}
\end{eqnarray}
where $\omega_{B}=eB/m$ is the cyclotron frequency associated
with the magnetic field oscillation amplitude, and the
dimensionless expansion parameter is given by the ratio
$\omega_{B}^2/\omega^2$. The effective potential
Eq.~(\ref{eq::VeffE}) for the field Eq.~(\ref{eq::Efield})
takes the form
\begin{eqnarray}
V_{\rm eff}&=&
{m\omega_B^2\over 16}\left[x^2\left(1
+{x^2-3z^2\over 3\Delta^2}\right)^2\right.
\nonumber\\
&+&\left.y^2\left(1+{y^2-3z^2\over 3\Delta^2}\right)^2
\right]\,.
\label{eq::Veff2d}
\end{eqnarray}
For the initial values  $z=\dot{z}=0$  the particle motion
described by Eq.~(\ref{eq::EOMtr}) remains  planar and the
potential Eq.~(\ref{eq::Veff2d}) plotted in Fig.~\ref{fig::3}c
provides charge confinement along $x$ and $y$ directions. An
example of the full and effective theory trajectories for the
motion in the $z=0$ plane is given in Fig.~\ref{fig::5}. An
unusual rectangular shape of the confining region is due to the
anharmonic terms in the potential.

The effective potential  Eq.~(\ref{eq::Veff2d}) vanishes along
the ``flat direction'' $x^2=y^2=3(z^2-\Delta^2)$ and, hence,
does not provide the confinement in all three dimensions. However,
by combining two  orthogonal phase-shifted oscillating
Ioffe-Pritchard fields we get the rotating magnetic field
\begin{equation}
{\bfm B}(t,{\bfm r})= {\bfm B}^{(x)}\cos(\omega t)-{\bfm
B}^{(y)}\sin(\omega t)\,,
\label{eq::Brot}
\end{equation}
where ${\bfm B}^{(x)}$ and ${\bfm B}^{(y)}$ are obtained by
rotating the field Eq.~(\ref{eq::Bfield}) into the $x$ and $y$
direction, respectively
\begin{eqnarray}
{\bfm B}^{(x)}&=&B\left(1+{y^2+z^2-2x^2\over 2\Delta^2},
{yx\over\Delta^2},{zx\over\Delta^2}\right)\,,
\nonumber\\
{\bfm B}^{(y)}&=&B\left({xy\over\Delta^2},
1+{z^2+x^2-2y^2\over 2\Delta^2},{zy\over\Delta^2}\right)\,.
\label{eq::Bfieldxy}
\end{eqnarray}
It is rather straightforward to derive the effective  potential
for such a field configuration with the result
\begin{eqnarray}
V_{\rm eff}&=&{m\omega_B^2\over 16}
\left[x^2\left(1+{x^2-3y^2\over 3\Delta^2}\right)^2
\!\!\!+y^2\left(1+{y^2-3x^2\over 3\Delta^2}\right)^2\right.
\nonumber\\
&+&\left.z^2\left(1+{z^2-3x^2\over 3\Delta^2}\right)^2
+z^2\left(1+{z^2-3y^2\over 3\Delta^2}\right)^2\right]\,.
\label{eq::Veff3d}
\end{eqnarray}
This potential does provide the dynamical magnetic confinement
in three dimensions and localizes  both positive and negative
electric charges  in a box-like region.  A typical trajectory
in the three-dimensional magnetic trap computed in the full and
effective theories is shown in Fig.~\ref{fig::6}. Though due
to the strong nonlinear effects the motion is rather irregular,
the effective theory accurately catches the qualitative
features of the exact solution.  In the limit $\Delta\to\infty$
the magnetic field Eq.~(\ref{eq::Brot}) becomes homogeneous and
the confining potential becomes harmonic, {\it i.e.} the
effective theory linearizes. Note that though
Eq.~(\ref{eq::EOMtr}) linearizes in this limit, the resulting
system is equivalent to a fourth-order ODE with the stability
chart different from the Mathieu equation.  This case has
been discussed in detail in \cite{Besharat:2025vbw}. Such a
rotating locally homogeneous magnetic field is realized in the
TOP traps for  magnetic moment of a neutral particle
\cite{Petrich:1995,Davis:1995}, where it plays an auxiliary
role to prevent the magnetic moment flip. The
inhomogeneous/nonlinear terms, however, could be mandatory for
some interesting  physical applications. Indeed, an oscillating
Ioffe-Pritchard field confines the magnetic moments
\cite{Besharat:2025vbw}. Thus, in principle, a trap with the
rapidly rotating inhomogeneous magnetic field
Eq.~(\ref{eq::Brot}) can be used to simultaneously confine the
positive and  negative charges as well as the neutral
particles with nonvanishing magnetic moments, which is
crucial   {\it e.g.} for the  antihydrogen production
\cite{Walz:1995,Andresen:2010}.

\vspace*{0mm}

\section{Summary}
\label{sec::sum}
In this work we have developed a systematic perturbative method
for the description of the nonlinear particle dynamics in  the
presence of rapidly oscillating forces, thereby providing a
general solution to the entire class of the problems introduced
by the original work \cite{Kapitza:1951}. The method is
optimized for the   calculation of the high-order terms in the
high-frequency expansion and offers a powerful tool for the
study of dynamical stabilization. It has been applied to a
number of real  physical systems and verified through the
numerical simulations. By analyzing the general all-order
structure of the high-frequency expansion we have obtained a
nonperturbative description of the stability properties for a
broad class of nonlinear systems, including the Kapitza
pendulum. Our results  can be used in a wide spectrum of
applications from the high-precision analysis and design of the
particle traps to the Floquet engineering of quantum
materials.\footnote{The  analysis of the high-frequency
effective theory quantization can be found in
\cite{Penin:2023dtn,Besharat:2025vbw}.}


\vspace*{4mm}
\noindent
{\bf Acknowledgments.}  The work of A.P. was supported in part
by NSERC and the Perimeter Institute for Theoretical Physics.
The work of A.B. is supported by NSERC.


\clearpage

\begin{widetext}

\vspace*{20mm}

\begin{figure}[h]
\begin{center}
\begin{tabular}{cc}
\hspace*{-3mm}\includegraphics[width=7cm]{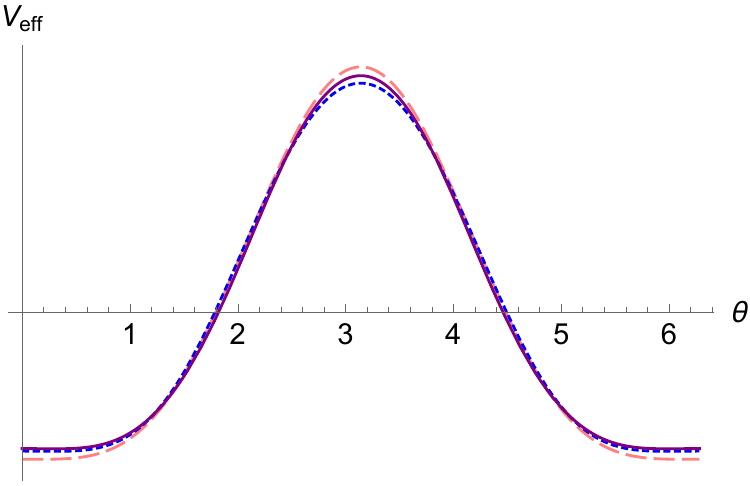}&
\hspace*{15mm}\includegraphics[width=7cm]{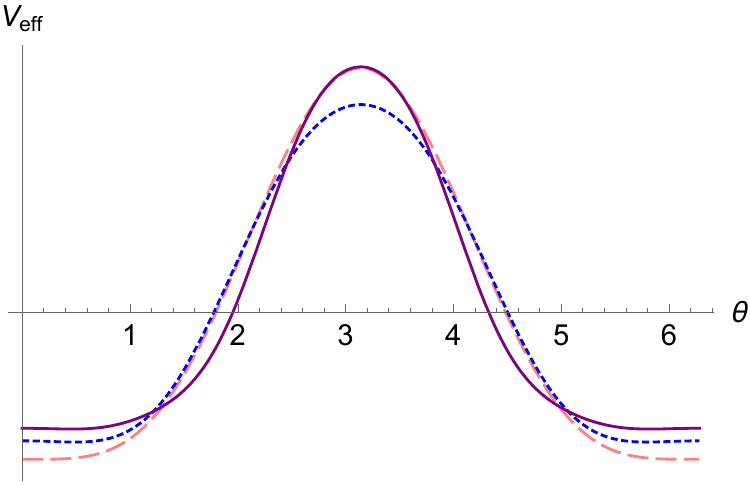}\\[5mm]
\hspace*{-3mm}(a)&
\hspace*{15mm}(b)\\
\end{tabular}
\end{center}
\caption{\label{fig::1} The convergence of the effective
potential for the horizontally driven Kapitza pendulum for the
near-critical value of the parameters  $\delta=\epsilon^2/2$
for (a)  $\epsilon=1/2$ and (b) $\epsilon=3/4$. The
long-dashed, short-dashed and solid lines represent the
leading, next-to-leading, and next-to-next-to-leading
approximations.}
\end{figure}

\vspace*{20mm}

\begin{figure}[h]
\begin{center}
\begin{tabular}{cc}
\hspace*{-3mm}\includegraphics[width=8cm]{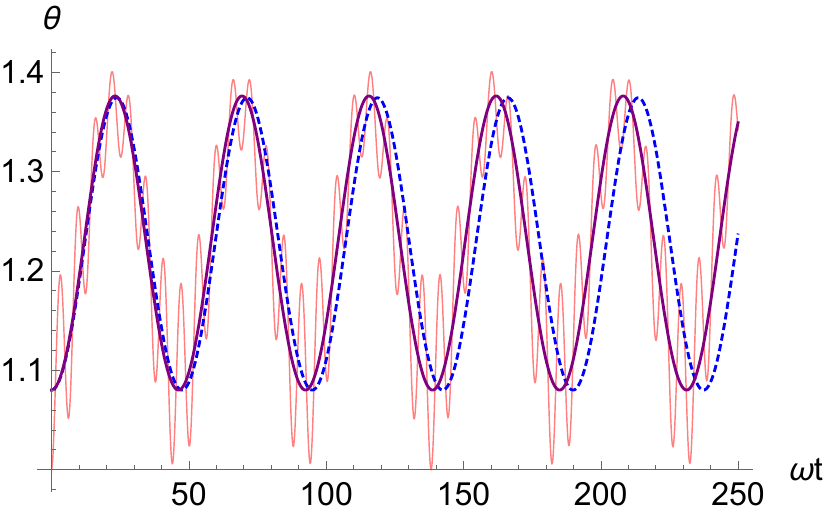}&
\hspace*{15mm}\includegraphics[width=8cm]{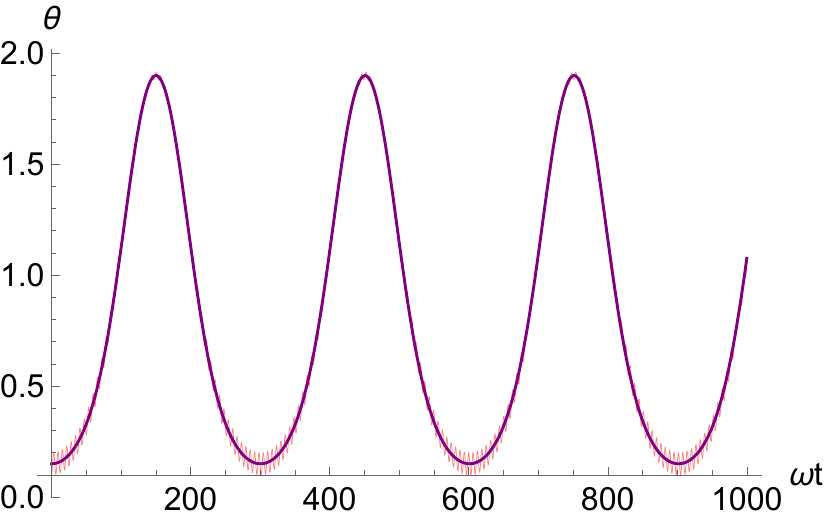}\\[5mm]
\hspace*{-3mm}(a)&
\hspace*{15mm}(b)\\
\end{tabular}
\end{center}
\caption{\label{fig::2} The solution of the exact equation of
motion for the planar pendulum with the horizontally oscillating
pivot (thin solid line) {\it vs} the leading-order (dashed
line), and the next-to-leading order (solid line)   effective
theory result for (a)  $\delta=\epsilon^2/6$, $\epsilon=1/5$
and (b) $\delta=\epsilon^2/6$, $\epsilon=1/20$. In the second
plot the two effective theory curves are indistinguishable.}
\end{figure}

\clearpage

\vspace*{20mm}

\begin{figure}[h]
\begin{center}
\begin{tabular}{ccc}
\hspace*{-3mm}\includegraphics[width=5cm]{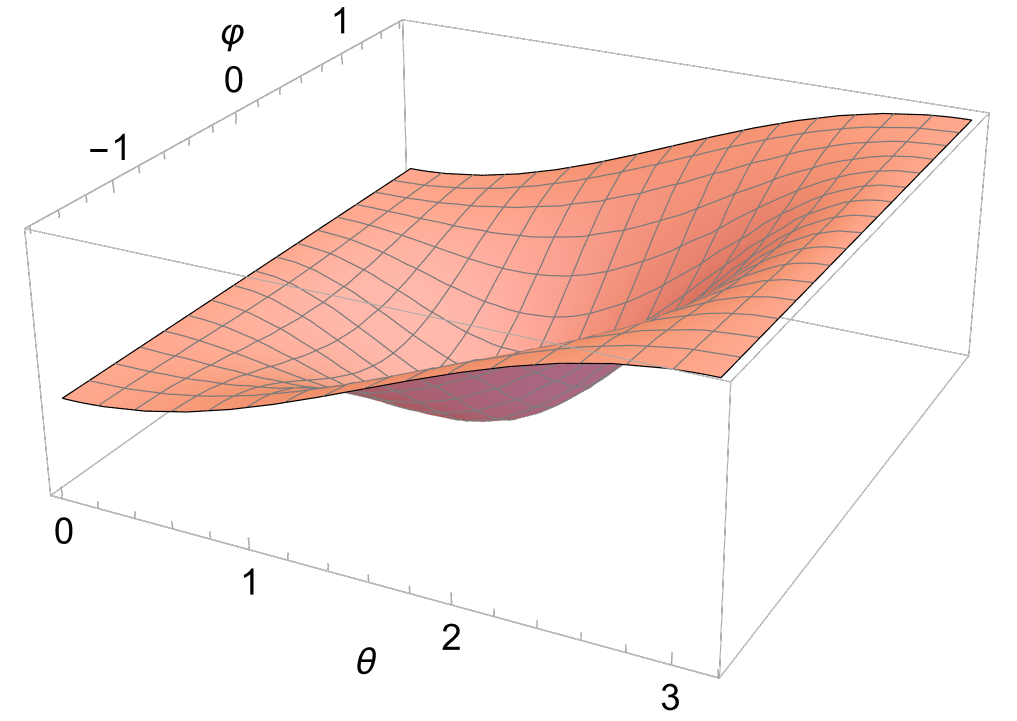}&
\hspace*{10mm}\includegraphics[width=5cm]{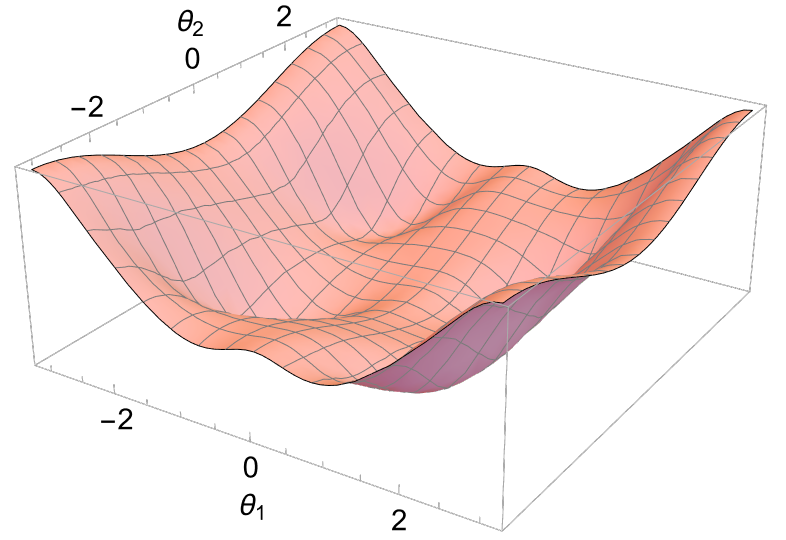}&
\hspace*{10mm}\includegraphics[width=5cm]{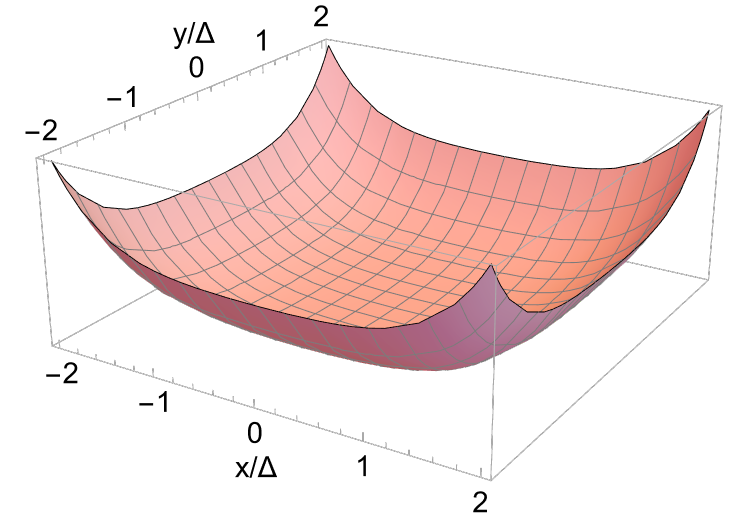}\\[5mm]
\hspace*{-3mm}(a)&
\hspace*{10mm}(b)&
\hspace*{10mm}(c)\\
\end{tabular}
\end{center}
\caption{\label{fig::3} The effective potential  for (a)
spherical  pendulum with $\delta=\epsilon^2/8$ (the symmetric
minimum at $\phi=\pi$ is not shown),  (b) planar double
pendulum with $\delta=\epsilon^2/6$, (c) oscillating magnetic
field of the Ioffe-Pritchard trap at $z=0$.}
\end{figure}

\vspace*{20mm}

\begin{figure}[h]
\begin{center}
\begin{tabular}{cc}
\hspace*{-3mm}\includegraphics[width=4cm]{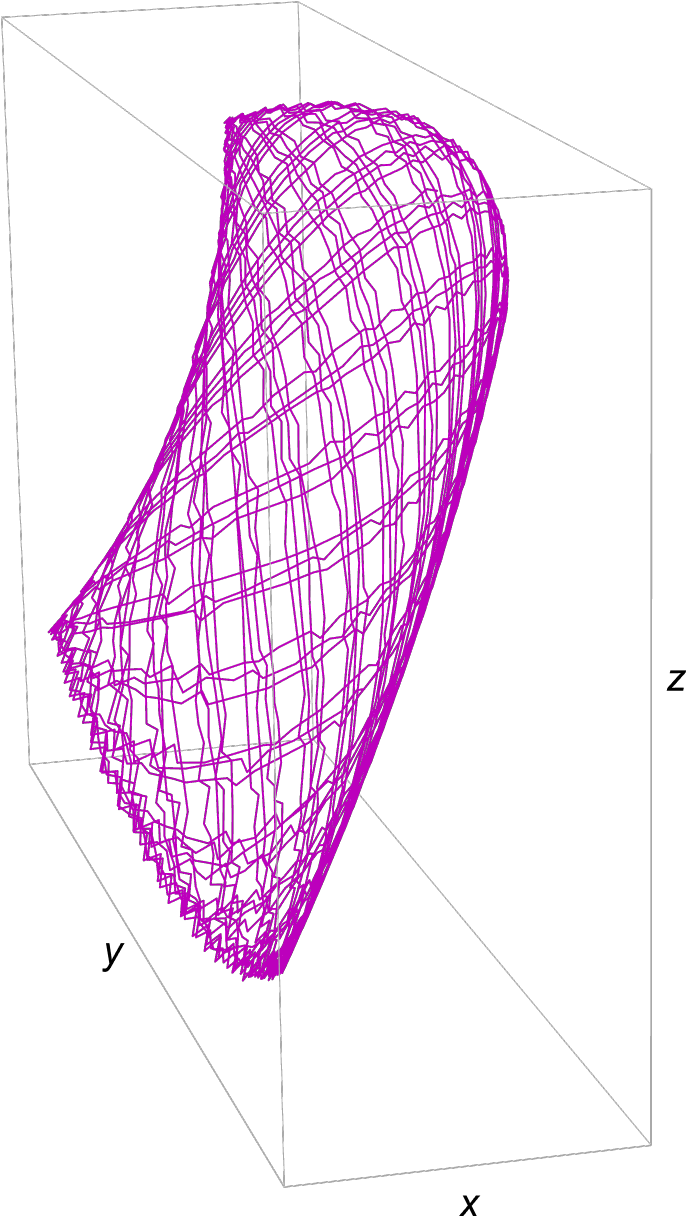}&
\hspace*{20mm}\includegraphics[width=4cm]{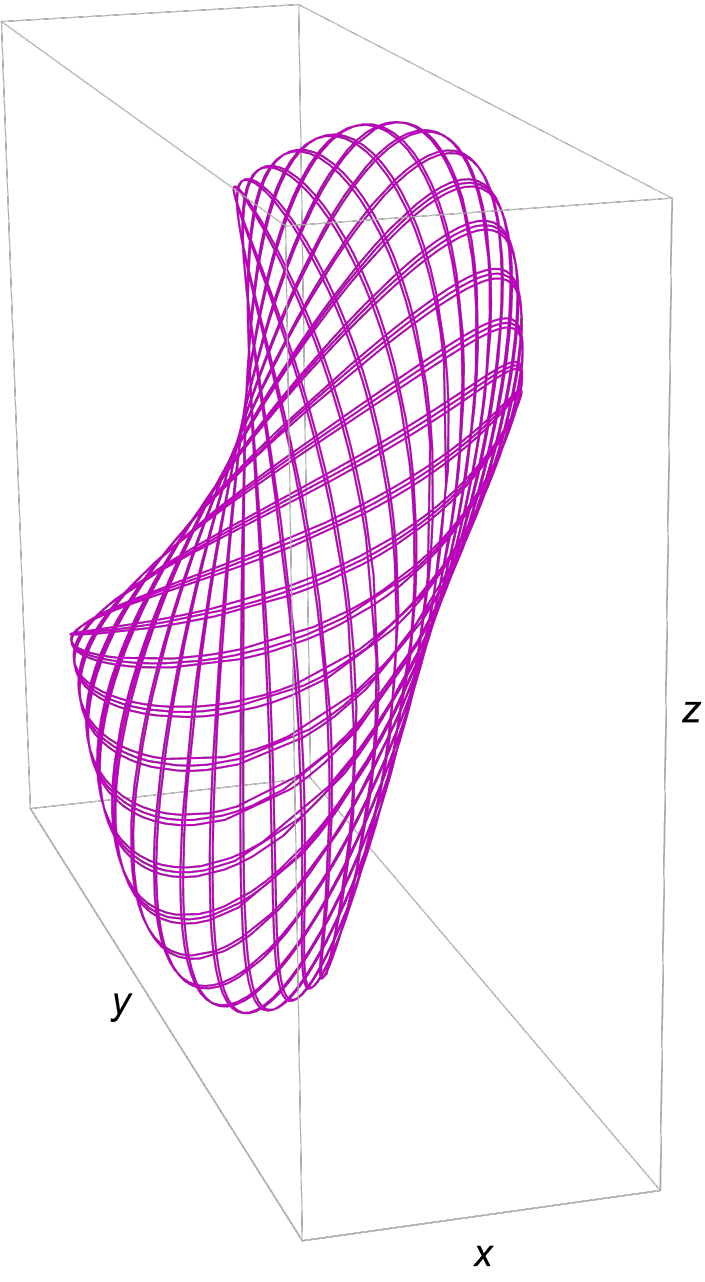}\\[5mm]
\hspace*{-3mm}(a)&
\hspace*{20mm}(b)\\
\end{tabular}
\end{center}
\caption{\label{fig::4} A  trajectory of the spherical pendulum
with the horizontally oscillating pivot for
$\delta=\epsilon^2/6$, $\epsilon=1/100$ obtained by solving (a)
the exact equations of motion and (b) the effective theory
equations of motion.}
\end{figure}

\clearpage

\vspace*{10mm}

\begin{figure}[h]
\begin{center}
\begin{tabular}{cc}
\hspace*{-3mm}\includegraphics[width=6.0cm]{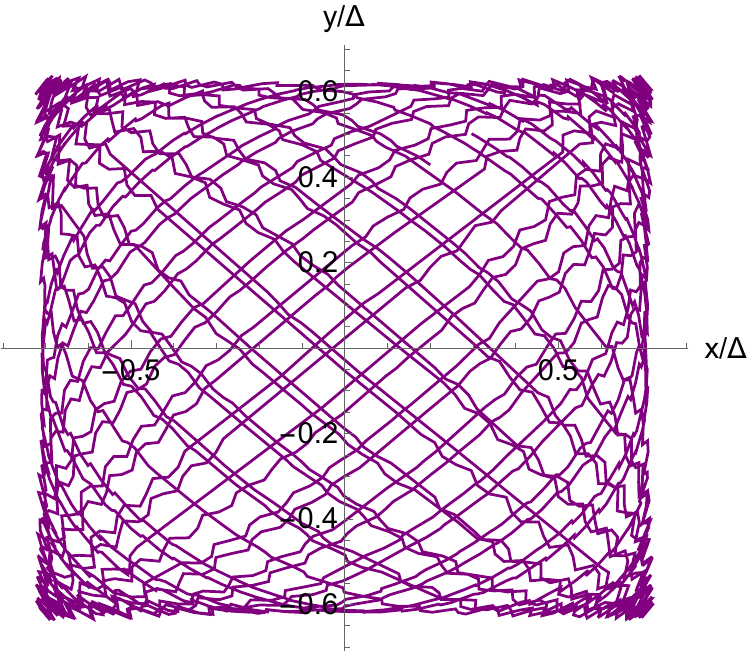}&
\hspace*{20mm}\includegraphics[width=6.0cm]{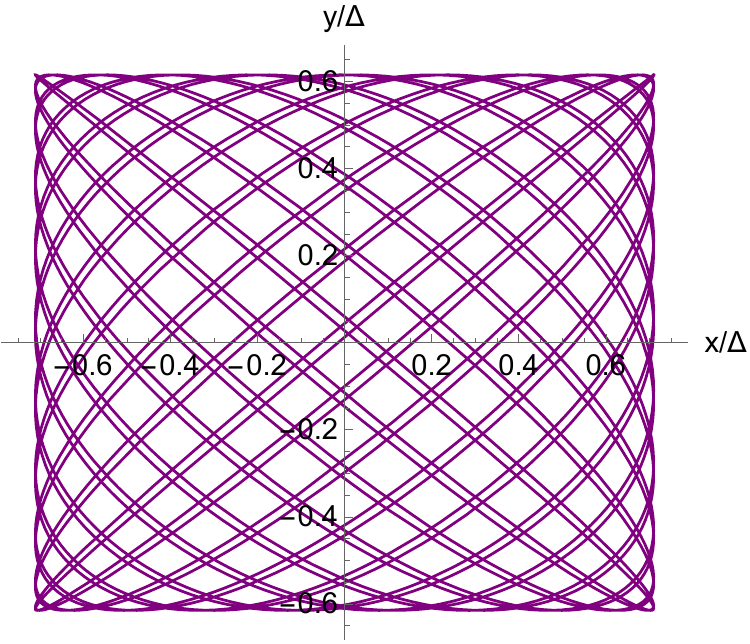}\\[5mm]
\hspace*{-3mm}(a)&
\hspace*{20mm}(b)\\
\end{tabular}
\end{center}
\caption{\label{fig::5} A  trajectory of a charged particle  in
the  oscillating magnetic field of the Ioffe-Pritchard trap for
$z=0$, $\omega_B/\omega=1/20$ obtained by solving (a) the exact
equations of motion and (b) the effective theory equations of
motion.}
\end{figure}

\vspace*{20mm}

\begin{figure}[h]
\begin{center}
\begin{tabular}{cc}
\hspace*{-3mm}\includegraphics[width=6.0cm]{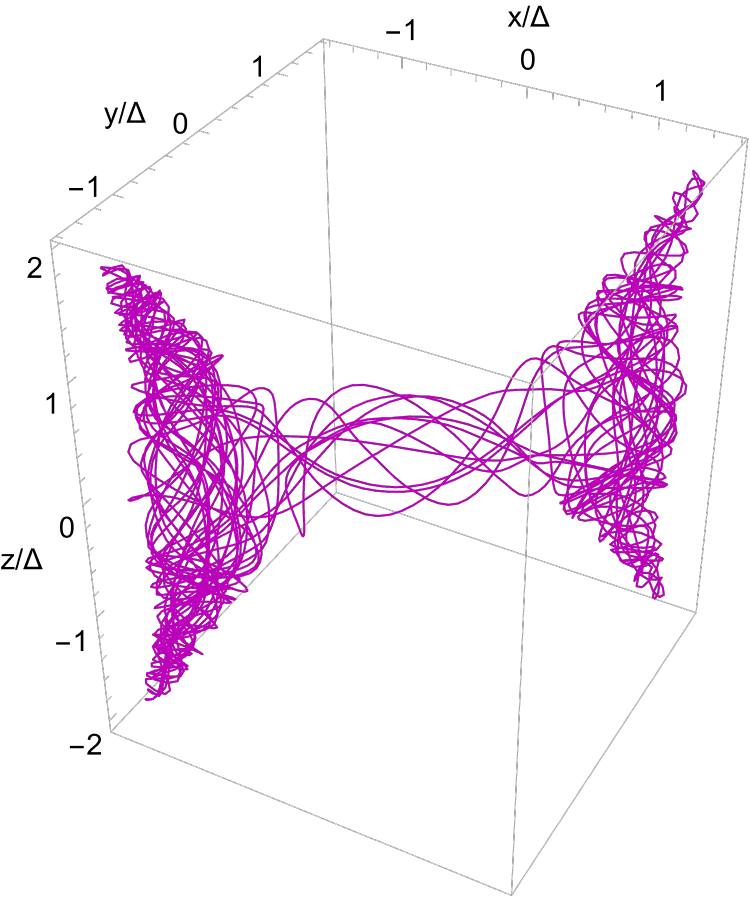}&
\hspace*{20mm}\includegraphics[width=6.0cm]{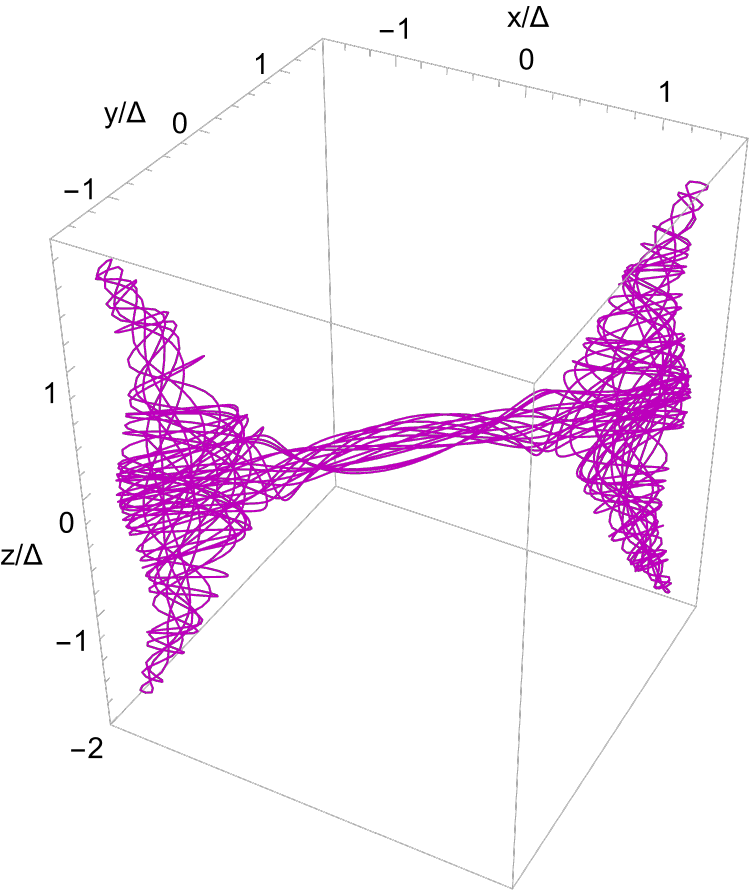}\\[5mm]
\hspace*{-3mm}(a)&
\hspace*{20mm}(b)\\
\end{tabular}
\end{center}
\caption{\label{fig::6} Same as Fig.~\ref{fig::5} but for the
three-dimensional magnetic trap and $\omega_B/\omega=1/50$.}
\end{figure}


\clearpage

\appendix

\section{${\cal O}(v^4)$ harmonic coefficients}
\label{sec::appA}

The non-vanishing harmonic coefficients in the next-to-leading
order of the high-frequency expansion read
\begin{eqnarray}
{\bfm c}_1^{(4)}&=&
-6v^{(1)}_iv^{(1)}_j\partial_i\partial_j{\bfm f}\,,
\nonumber\\
{\bfm s}_1^{(4)}&=&-2v^{(3)}_i \partial_i{\bfm f}
-4v^{(1)}_i(\partial_if_j)\partial_j{\bfm g}
-12g_iv^{(1)}_j\partial_i\partial_j{\bfm f}
-2f_iv^{(1)}_j\partial_i\partial_j{\bfm g}
-4v^{(1)}_i(\partial_ig_j)\partial_j{\bfm f}
-2v^{(1)}_if_j(\partial_if_k)
\partial_j\partial_k{\bfm f}
\nonumber\\
&-&{49\over 8}v^{(1)}_if_j(\partial_jf_k)
\partial_i\partial_k{\bfm f}
+4v^{(1)}_iv^{(1)}_jv^{(1)}_k
\partial_i\partial_j\partial_k{\bfm f}
-{3\over 4}v^{(1)}_if_jf_k
\partial_i\partial_j\partial_k{\bfm f}
-{35\over 16}v^{(1)}_if_j(\partial_i\partial_jf_k)
\partial_k{\bfm f}
\nonumber\\
&-&{37\over 16}v^{(1)}_i(\partial_if_j)(\partial_jf_k)
\partial_k{\bfm f}\,,
\nonumber\\
{\bfm c}_2^{(4)}&=&{1\over 32} f_i(\partial_i f_j)
\partial_j{\bfm g}
+{3\over 32}g_if_j\partial_i\partial_j{\bfm f}
+{1\over 32}f_if_j\partial_i\partial_j{\bfm g}
+{23\over 32}g_i(\partial_i f_j)\partial_j{\bfm f}
+{1\over 8}f_i(\partial_ig_j)\partial_j{\bfm f}
+{1\over 16}f_if_j(\partial_jf_k)
\partial_i\partial_k{\bfm f}
\nonumber\\
&-&{7\over 16}v^{(1)}_iv^{(1)}_j(\partial_if_k)
\partial_j\partial_k{\bfm f}
+{1\over 48}f_if_jf_k
\partial_i\partial_j\partial_k{\bfm f}
-{3\over 32}v^{(1)}_iv^{(1)}_jf_k
\partial_i\partial_j\partial_k{\bfm f}
+{7\over 144}f_if_j(\partial_i\partial_jf_k)
\partial_k{\bfm f}
\nonumber\\
&-&{23\over 32}v^{(1)}_iv^{(1)}_j
(\partial_i\partial_jf_k)\partial_k{\bfm f}
+{53\over 144}f_i(\partial_if_j)(\partial_jf_k)
\partial_k{\bfm f}\,,
\nonumber\\
{\bfm s}_2^{(4)}&=&-{1\over 8}v^{(2)}_if_j
\partial_i\partial_i {\bfm f}
-{3\over 8}v^{(2)}_i(\partial_i f_j)
\partial_j{\bfm f}
\,,
\nonumber\\
{\bfm s}_3^{(4)}&=&-{2\over 27}v^{(1)}_if_j
(\partial_if_k)\partial_j\partial_k{\bfm f}
-{1\over 216}v^{(1)}_if_j
(\partial_jf_k)\partial_i\partial_k{\bfm f}
-{1\over 108}v^{(1)}_if_jf_k
\partial_i\partial_j\partial_k{\bfm f}
-{5\over 432}v^{(1)}_if_j(\partial_i\partial_jf_k)
\partial_k{\bfm f}
\nonumber\\
&-&{11\over 432}v^{(1)}_i(\partial_if_j)(\partial_jf_k)
\partial_k{\bfm f}\, ,
\nonumber\\
{\bfm c}_4^{(4)}&=&\frac{1}{2304}f_i
f_j(\partial_{i}\partial_{j}f_k)
\partial_k {\bfm f}+\frac{1}{4608}f_i(\partial_i f_j)
(\partial_j f_k)\partial_k {\bfm f}
+\frac{1}{512}f_i f_j (\partial_j f_k) \partial_i
\partial_k {\bfm f}+\frac{1}{768} f_i f_j
f_k \partial_i \partial_j \partial_k {\bfm f}\,.
\label{eq::v4}
\end{eqnarray}

\clearpage

\section{${\cal O}(v^6)$ effective force}
\label{sec::appB}
The effective force  in the next-to-next-to-leading
order of the high-frequency expansion reads
\begin{align}
{\bfm {\cal F}}^{(6)}_{\rm eff} &=
\frac{5}{2}v^{(1)}_{i}v^{(1)}_{j}v^{(1)}_{k}v^{(1)}_{l}
(\partial_{i}\partial_{j}\partial_{k}\partial_{l}f_{m})
\partial_{m}\bm{f}
+\frac{3}{2} ({ {\cal F}}^{(4)}_{\rm eff})_i (\partial_{i}f_{j})\partial_{j}\bm{f}
-\frac{3}{2}\,v^{(2)}_{\,i}v^{(2)}_{\,j}(\partial_{i}\partial_{j}f_{k})\,\partial_{k}\bm{f}
-3\,v^{(1)}_{i}v^{(3)}_{j}(\partial_{i}\partial_{j}f_{k})\partial_{k}\bm{f}
\nonumber\\
&\quad +v^{(1)}_{i}v^{(1)}_{j}\bigg[-\frac{3}{2}f_{k}(\partial_{i}\partial_{j}\partial_{k}g_{l})\partial_{l}\bm{f}
-\frac{3}{2}f_{k}(\partial_{i}\partial_{j}f_{l})\partial_{l}\partial_{k}\bm{g}
-5(\partial_{i}f_{k})(\partial_{j}\partial_{k}g_{l})\partial_{l}\bm{f}
\nonumber\\
&\quad+(\partial_{i}f_{k})(\partial_{j}f_{l})\partial_{k}\partial_{l}\bm{g}
-\frac{5}{2}(\partial_{i}\partial_{j}g_{k})(\partial_{k}f_{l})\partial_{l}\bm{f}
-5 (\partial_{i}\partial_{j}f_{k})(\partial_{k}g_{l})\partial_{l}\bm{f}
\nonumber\\
&\quad-10(\partial_{i}g_{k})(\partial_{j}\partial_{k}f_{l})\partial_{l}\bm{f}
-15g_{k}(\partial_{i}\partial_{j}\partial_{k}f_{l})\partial_{l}\bm{f}
-\frac{9}{16}f_{k}f_{l}
(\partial_{i}\partial_{j}\partial_{k}\partial_{l}f_{m})
\partial_{m}\bm{f}
\nonumber\\
&\quad-\frac{11}{4}f_{k}(\partial_{j}f_{l})(\partial_{i}\partial_{k}\partial_{l}f_{m})\partial_{m}\bm{f}
-\frac{243}{32}f_{k}(\partial_{k}f_{l})(\partial_{j}\partial_{i}\partial_{l}f_{m})\partial_{m}\bm{f}
-\frac{183}{128}f_{k}
(\partial_{i}\partial_{j}\partial_{k}f_{l})(\partial_{l}f_{m})
\partial_{m}\bm{f}
\nonumber\\
&\quad-\frac{3}{128}f_{k}f_{l}
(\partial_{l}\partial_{i}\partial_{j}f_{m})
\partial_{k}\partial_{m}\bm{f}
+\frac{1}{4}f_{k}(\partial_{j}f_{l})(\partial_{i}f_{m})\partial_{k}\partial_{l}\partial_{m}\bm{f}
-\frac{9}{16}f_{k}f_{l}(\partial_{i}\partial_{j}f_{m})\partial_{k}\partial_{l}\partial_{m}\bm{f}
\nonumber\\
&\quad
-\frac{21}{4}f_{k}
(\partial_{i}\partial_{k}f_{l})(\partial_{j}\partial_{l}f_{m})\partial_{m}\bm{f}
-\frac{11}{4}f_{k}
(\partial_{j}\partial_{i}f_{l})(\partial_{k}\partial_{l}f_{m})
\partial_{m}\bm{f}-\frac{11}{8}(\partial_{i}f_{k})(\partial_{j}f_{l})(\partial_{k}\partial_{l}f_{m})\partial_{m}\bm{f}
\nonumber\\
&\quad
-\frac{219}{128}
(\partial_{i}\partial_{j}f_{k})(\partial_{k}f_{l})(\partial_{l}f_{m})
\partial_{m}\bm{f}
-\frac{23}{128}f_{k}
(\partial_{j}\partial_{i}f_{l})(\partial_{l}f_{m})
\partial_{k}\partial_{m}\bm{f}-\frac{7}{64}f_{k}
(\partial_{j}f_{l})(\partial_{i}\partial_{l}f_{m})
\partial_{k}\partial_{m}\bm{f}
\nonumber\\
&\quad
+\frac{1}{16}f_{k}
(\partial_{j}f_{l})(\partial_{i}\partial_{k}f_{m})
\partial_{l}\partial_{m}\bm{f}
-\frac{3}{32}f_{k}
(\partial_{k}f_{l})(\partial_{j}\partial_{i}f_{m})
\partial_{l}\partial_{m}\bm{f}-\frac{43}{8}(\partial_{i}f_{k})(\partial_{k}f_{l})(\partial_{j}\partial_{l}f_{m})\partial_{m}\bm{f}
\nonumber\\
&\quad
-\frac{195}{64}(\partial_{i}f_{k})(\partial_{j}\partial_{k}f_{l})(\partial_{l}f_{m})\partial_{m}\bm{f}
+\frac{3}{16}(\partial_{i}f_{k})(\partial_{j}f_{l})(\partial_{k}f_{m})\partial_{l}\partial_{m}\bm{f}\bigg]\nonumber\\
&\quad+\frac{1}{64}f_{i}f_{j}f_{k}f_{l}\partial_{i}\partial_{j}\partial_{k}\partial_{l}\bm{g}
+\frac{1}{64}f_{i}f_{j}f_{k}\,(\partial_{i}f_{l})\partial_{j}\partial_{k}\partial_{l}\bm{g}
+\frac{1}{16}\,f_{i}f_{j}f_{k}(\partial_{i}\partial_{j}\partial_{k}g_{l})\partial_{l}\bm{f}
+\frac{3}{2}f_{i}g_{j}(\partial_{j}f_{k})\partial_{i}\partial_{k}\bm{g}
\nonumber\\
&\quad+\frac{1}{2} f_{i}f_{j} (\partial_{i}g_{k})\partial_{j}\partial_{k}\bm{g}
+\frac{3}{2}f_{i}g_{j}(\partial_{i}\partial_{j}g_{k})\partial_{k}\bm{f}
+\frac{25}{32}f_{i}f_{j}(\partial_{i}f_{k})(\partial_{k}f_{l})\partial_{j}\partial_{l}\bm{g}
+\frac{25}{32}f_{i}f_{j}(\partial_{i}f_{k})(\partial_{j}\partial_{k}g_{l})\partial_{l}\bm{f}
\nonumber\\
&\quad+\frac{1}{256}f_{i}f_{j}(\partial_{i}f_{k})(\partial_{j}f_{l})\partial_{k}\partial_{l}\bm{g}
+\frac{3}{16}\,f_{i}f_{j}f_{k}(\partial_{i}\partial_{j}f_{l})\partial_{k}\partial_{l}\bm{g}
+\frac{1}{64}f_{i}f_{j}(\partial_{i}\partial_{j}g_{k})(\partial_{k}f_{l})\partial_{l}\bm{f}
+\frac{1}{64}f_{i}f_{j}f_{k}(\partial_{j}\partial_{k}g_{l})\partial_{i}\partial_{l}\bm{f}
\nonumber\\
&\quad+5 g_{i}(\partial_{i}f_{j})(\partial_{j}g_{k})\partial_{k}\bm{f}
+\frac{5}{2}\,g_{i} (\partial_{i}g_{j})(\partial_{j}f_{k}) \partial_{k}\bm{f}
+\frac{1}{2} f_{i} (\partial_{i}g_{j}) (\partial_{j}g_{k}) \partial_{k}\bm{f}
+\frac{81}{32}\,f_{i} (\partial_{i}f_{j}) (\partial_{j}f_{k}) (\partial_{k}g_{l}) \partial_{l}\bm{f}
\nonumber\\
&\quad+\frac{161}{128}f_{i}(\partial_{i}f_{j})(\partial_{j}g_{k})(\partial_{k}f_{l})\partial_{l}\bm{f}
+\frac{1}{128}f_{i}f_{j}(\partial_{i}f_{k})(\partial_{k}g_{l})\partial_{l}\partial_{j}\bm{f}
+\frac{3}{8}f_{i}f_{j}(\partial_{i}g_{k})(\partial_{j}\partial_{k}f_{l})\partial_{l}\bm{f}
+\frac{1}{32}\,f_{i}f_{j}(\partial_{i}f_{k})(\partial_{j}g_{l})\partial_{k}\partial_{l}\bm{f}
\nonumber\\
&\quad+\frac{3}{16}f_{i}f_{j}f_{k} (\partial_{i}g_{l})\partial_{j}\partial_{k}\partial_{l}\bm{f}
+\frac{1}{32}f_{i}(\partial_{i}g_{j})(\partial_{j}f_{k})(\partial_{k}f_{l})\partial_{l}\bm{f}
+\frac{1}{32} f_{i}f_{j}(\partial_{i}g_{k})(\partial_{k}f_{l})\partial_{j}\partial_{l}\bm{f}
+\frac{3}{16}f_{i}f_{j}(\partial_{i}\partial_{j}f_{k})(\partial_{k}g_{l})\partial_{l}\bm{f}
\nonumber\\
&\quad+\frac{15}{2}\,g_{i}g_{j}(\partial_{i}\partial_{j}f_{k})\partial_{k}\bm{f}
+\frac{23}{128}\,f_{i}g_{j}(\partial_{j}f_{k})(\partial_{k}f_{l})\partial_{l}\partial_{i}\bm{f}
+\frac{11}{4}f_{i}g_{j}(\partial_{j}f_{k})(\partial_{k}\partial_{i}f_{l})\partial_{l}\bm{f}
+\frac{243}{32}f_{i}g_{j}(\partial_{i}f_{k})(\partial_{k}\partial_{j}f_{l})\partial_{l}\bm{f}
\nonumber\\
&\quad+\frac{9}{16}f_{i}f_{j}g_{k}(\partial_{k}f_{l})\partial_{i}\partial_{j}\partial_{l}\bm{f}
+\frac{3}{32}f_{i}g_{j}(\partial_{i}f_{k})(\partial_{j}f_{l})\partial_{k}\partial_{l}\bm{f}
+\frac{9}{16}f_{i}f_{j}g_{k}(\partial_{i}\partial_{j}\partial_{k}f_{l})\partial_{l}\bm{f}
+\frac{183}{128}f_{i}g_{j}(\partial_{i}\partial_{j}f_{k})(\partial_{k}f_{l})\partial_{l}\bm{f}
\nonumber\\
&\quad+\frac{219}{128}g_{i}(\partial_{i}f_{j})(\partial_{j}f_{k})(\partial_{k}f_{l})\partial_{l}\bm{f}
+\frac{3}{128}\,f_{i}f_{j}g_{k}(\partial_{j}\partial_{k}f_{l})\partial_{i}\partial_{l}\bm{f}
+\frac{1}{384}f_{i}f_{j}f_{k}f_{l}f_{m}
\partial_{i}\partial_{j}\partial_{k}\partial_{l}\partial_{m}\bm{f}
\nonumber\\
&\quad+\frac{1}{192}f_{i}f_{j}f_{k}f_{l}(\partial_{i}f_{m})
\partial_{k}\partial_{j}\partial_{l}\partial_{m}\bm{f}
+\frac{5}{384}\,f_{i}f_{j}f_{k}f_{l}
(\partial_{i}\partial_{j}\partial_{k}\partial_{l}f_{m})
\partial_{m}\bm{f}
+\frac{169}{576}f_{i}f_{j}f_{k}(\partial_{i}f_{l})(\partial_{l}f_{m})
\partial_{j}\partial_{k}\partial_{m}\bm{f}
\nonumber\\
&\quad+\frac{1}{512}f_{i}f_{j}f_{k}(\partial_{i}f_{l})(\partial_{j}f_{m})
\partial_{m}\partial_{l}\partial_{k}\bm{f}
+\frac{41}{576}f_{i}f_{j}f_{k}f_{l}(\partial_{i}\partial_{j}f_{m})
\partial_{k}\partial_{l}\partial_{m}\bm{f}
+\frac{19}{64}\,f_{i}f_{j}f_{k}(\partial_{i}f_{l})
(\partial_{j}\partial_{k}\partial_{l}f_{m})\partial_{m}\bm{f}
\nonumber\\
&\quad+\frac{1}{192}f_{i}f_{j}f_{k}
(\partial_{i}\partial_{j}\partial_{k}f_{l})(\partial_{l}f_{m})
\partial_{m}\bm{f}
+\frac{1}{192}f_{i}f_{j}f_{k}f_{l}
(\partial_{j}\partial_{k}\partial_{l}f_{m})\partial_{i}\partial_{m}\bm{f}
+\frac{53}{576}f_{i}f_{j}(\partial_{i}f_{k})(\partial_{k}f_{l})
(\partial_{l}f_{m})\partial_{j}\partial_{m}\bm{f}
\nonumber\\
&\quad+\frac{403}{288}f_{i}f_{j}(\partial_{i}f_{k})(\partial_{k}f_{l})
(\partial_{j}\partial_{l}f_{m})\partial_{m}\bm{f}
+\frac{23}{32}f_{i}f_{j}(\partial_{i}f_{k})
(\partial_{j}\partial_{k}f_{l})(\partial_{l}f_{m})\partial_{m}\bm{f}
+\frac{113}{2304}f_{i}f_{j}\,(\partial_{i}f_{k})(\partial_{j}f_{l})
(\partial_{k}f_{m})\partial_{l}\partial_{m}\bm{f}
\nonumber\\
&\quad+\frac{985}{512}\,f_{i}f_{j}(\partial_{i}f_{k})(\partial_{j}f_{l})
(\partial_{k}\partial_{l}f_{m})\partial_{m}\bm{f}
+\frac{7}{576}f_{i}f_{j}
(\partial_{i}\partial_{j}f_{k})(\partial_{k}f_{l})(\partial_{l}f_{m})
\partial_{m}\bm{f}
+\frac{7}{576}\,f_{i}f_{j}f_{k}
(\partial_{k}\partial_{j}f_{l})
(\partial_{l}f_{m})\partial_{i}\partial_{m}\bm{f}
\nonumber\\
&\quad+\frac{41}{288}f_{i}f_{j}f_{k}
(\partial_{j}\partial_{k}f_{l})(\partial_{i}\partial_{l}f_{m})
\partial_{m}\bm{f}
+\frac{1}{64}f_{i}f_{j}f_{k}
(\partial_{i}f_{l})(\partial_{k}\partial_{l}f_{m})
(\partial_{j}\partial_{m}\bm{f})
+\frac{7}{576}f_{i}f_{j}f_{k}
(\partial_{i}f_{l})(\partial_{j}\partial_{k}f_{m})
(\partial_{l}\partial_{m}\bm{f})
\nonumber\\
&\quad+\frac{247}{288}f_{i}
(\partial_{i}f_{j})
(\partial_{j}f_{k})(\partial_{k}f_{l})(\partial_{l}f_{m})
\partial_{m}\bm{f}\,.
\label{eq::v6}
\end{align}

\clearpage

\section{Derivation of the ${\cal O}(v^6)$ effective Lagrangian}
\label{sec::appC}
Let us outline the derivation of the Lagrangian for the
velocity-dependent force Eq.~(\ref{eq::v6}). The general idea
is to list all possible tensor structures with the total number
of spatial derivatives  equal 4, the total power of the
velocity equal $2n$, the total power of the $\bfm f$ field
equal $2m$, and the total power of  the  $\bfm g$ field equal
$3-m-n$, with $n=0,\,1$, $m=1,\,2,\,3$, $m+n\le 3$. A large
number of the allowed tensor structures is significantly
reduced by eliminating the terms incompatible with the known
one-dimensional expression. The Lagrangian is taken as the
linear combination of these structures with arbitrary
coefficients, which are then determined by comparing the
corresponding  Euler-Lagrange equation to $\dot{\bfm
v}^{(6)}+{\bfm {\cal F}}^{(6)}_{\rm eff} =0$.   This can be
done for all the terms in Eq.~(\ref{eq::v6}) except the first
one
\begin{equation}
\frac{5}{2}v_{i}v_{j}v_{k}v_{l}
(\partial_{i}\partial_{j}\partial_{k}\partial_{l}f_{m})
\partial_{m}\bm{f}\,,
\label{eq::Fvvvv}
\end{equation}
which includes the forth power of the velocity.  The
relevant structure in the Lagrangian is of the form
\begin{equation}
{5\over 12}v_iv_jv_kv_l
\left[2(\partial_i\partial_j\partial_k{\bfm f})
\partial_l{\bfm f}-(\partial_i\partial_j{\bfm f})
\partial_k\partial_l{\bfm f}\right]\,.
\label{eq::Lvvvv}
\end{equation}
It gives the following contribution to the Euler-Lagrange
equation
\begin{eqnarray}
&&(v_l\partial_l)^2\left\{{5\over 3}v_iv_j
\left[(\partial_i\partial_jf_k){\bfm \partial}f_k
-(\partial_i{\bfm \partial}f_k)\partial_jf_k\right]\right\}
+\frac{5}{2}v_{i}v_{j}v_{k}v_{l}
(\partial_{i}\partial_{j}\partial_{k}\partial_{l}f_{m})
\partial_{m}\bm{f}+\ldots=0\,,
\label{eq::ELeqvvvv}
\end{eqnarray}
where the second term exactly reproduces Eq.~(\ref{eq::Fvvvv})
and the ellipsis stands for the terms which, after the use of
the leading order equation of motion,  include at most the
second power of the velocity. These terms can be subtracted
from the rest of Eq.~(\ref{eq::v6}) and the  part of the
Lagrangian Eq.~(\ref{eq::Leffnnlo}) corresponding to the
resulting expression can be obtained as described in the
beginning of this section. The first term in
Eq.~(\ref{eq::ELeqvvvv}) cannot be accommodated in this way.
However, since for the slow variables
$(v_l\partial_l)^2={d^2/dt^2}$, this is the second order time
derivative which can be absorbed into the acceleration term in
the equation of motion by the coordinate transformation
\begin{equation}
\dot{\bfm
v}+{d^2\over dt^2}\left\{{5\over 3}v_iv_j
\left[(\partial_i\partial_jf_k){\bfm \partial}f_k
-(\partial_i{\bfm \partial}f_k)\partial_jf_k\right]\right\}\to
\dot{\bfm
v}'\,,
\label{eq::vp}
\end{equation}
where ${\bfm v}'=\dot{\bfm r}'$ and ${\bfm r}'$ is given by
Eq.~(\ref{eq::rp}). Then, we need to express the effective
Lagrangian in terms of the new variable. Since the correction
term in Eq.~(\ref{eq::rp}) is ${\cal O}(v^4)$, to get the
${\cal O}(v^6)$ approximation the coordinate transformation
should be taken into account in the leading ${\cal O}(v^2)$
part of the effective Lagrangian only. This is done by
substituting
\begin{equation}
{\bfm r}={\bfm r}'-{5\over 3}v'_iv'_j
\left[(\partial_i\partial_jf_k({\bfm r}')){\bfm \partial}f_k({\bfm r}')
-(\partial_i{\bfm \partial}f_k({\bfm r}'))\partial_jf_k({\bfm r}')\right]\,
\label{eq::r}
\end{equation}
into Eq.~(\ref{eq::Leffv2}) and expanding the resulting
expression in Taylor series about ${\bfm r}'$, which generates
an additional next-to-next-to-leading  contribution to
Eq.~(\ref{eq::Leffnnlo}). Note that in Eq.~(\ref{eq::r}) we
have replaced ${\bfm r}$ with ${\bfm r}'$ in the correction
term, which is valid through ${\cal O}(v^6)$.

\end{widetext}

\end{document}